
\magnification=1200
\hoffset-.6truecm\voffset-.8truecm
\hsize=17truecm\vsize=25truecm
%
\newbox\leftpage \newdimen\fullhsize \newdimen\hstitle \newdimen\hsbody
\tolerance=1000\hfuzz=2pt
\catcode`\@=11 
\newcount\yearltd\yearltd=\year\advance\yearltd by -1900
%

\def\draftmode{\message{ DRAFTMODE }\def\draftdate{{\rm preliminary draft:
\number\month/\number\day/\number\yearltd\ \ \hourmin}}%
\headline={\hfil\draftdate}\writelabels\baselineskip=20pt plus 2pt minus 2pt
 {\count255=\time\divide\count255 by 60 \xdef\hourmin{\number\count255}
  \multiply\count255 by-60\advance\count255 by\time
  \xdef\hourmin{\hourmin:\ifnum\count255<10 0\fi\the\count255}}}
\def\nolabels{\def\wrlabeL##1{}\def\eqlabeL##1{}\def\reflabeL##1{}}
\def\writelabels{\def\wrlabeL##1{\leavevmode\vadjust{\rlap{\smash%
{\line{{\escapechar=` \hfill\rlap{\sevenrm\hskip.03in\string##1}}}}}}}%
\def\eqlabeL##1{{\escapechar-1\rlap{\sevenrm\hskip.05in\string##1}}}%
\def\reflabeL##1{\noexpand\llap{\noexpand\sevenrm\string\string\string##1}}}
\nolabels
%
\global\newcount\secno \global\secno=0
\global\newcount\meqno \global\meqno=1
\def\newsec#1{\global\advance\secno by1\message{(\the\secno. #1)}
\global\subsecno=0\eqnres@t\noindent{\bf\the\secno. #1}
\writetoca{{\secsym} {#1}}\par\nobreak\medskip\nobreak}
\def\eqnres@t{\xdef\secsym{\the\secno.}\global\meqno=1\bigbreak\bigskip}
\def\sequentialequations{\def\eqnres@t{\bigbreak}}\xdef\secsym{}
\global\newcount\subsecno \global\subsecno=0
\def\subsec#1{\global\advance\subsecno by1\message{(\secsym\the\subsecno. #1)}
\ifnum\lastpenalty>9000\else\bigbreak\fi
\noindent{\it\secsym\the\subsecno. #1}\writetoca{\string\quad
{\secsym\the\subsecno.} {#1}}\par\nobreak\medskip\nobreak}
\def\appendix#1#2{\global\meqno=1\global\subsecno=0\xdef\secsym{\hbox{#1.}}
\bigbreak\bigskip\noindent{\bf Appendix #1. #2}\message{(#1. #2)}
\writetoca{Appendix {#1.} {#2}}\par\nobreak\medskip\nobreak}
%
%
\def\eqnn#1{\xdef #1{(\secsym\the\meqno)}\writedef{#1\leftbracket#1}%
\global\advance\meqno by1\wrlabeL#1}
\def\eqna#1{\xdef #1##1{\hbox{$(\secsym\the\meqno##1)$}}
\writedef{#1\numbersign1\leftbracket#1{\numbersign1}}%
\global\advance\meqno by1\wrlabeL{#1$\{\}$}}
\def\eqn#1#2{\xdef #1{(\secsym\the\meqno)}\writedef{#1\leftbracket#1}%
\global\advance\meqno by1$$#2\eqno#1\eqlabeL#1$$}
%
\newskip\footskip\footskip14pt plus 1pt minus 1pt 
%
%

%
%
\global\newcount\refno \global\refno=1
\newwrite\rfile
\def\ref{$^{\the\refno)}$\nref}
\def\nref#1{\xdef#1{[\the\refno]}\writedef{#1\leftbracket#1}%
\ifnum\refno=1\immediate\openout\rfile=refs.tmp\fi
\global\advance\refno by1\chardef\wfile=\rfile\immediate
\write\rfile{\noexpand\item{#1\ }\reflabeL{#1\hskip.31in}\pctsign}\findarg}
\def\findarg#1#{\begingroup\obeylines\newlinechar=`\^^M\pass@rg}
{\obeylines\gdef\pass@rg#1{\writ@line\relax #1^^M\hbox{}^^M}%
\gdef\writ@line#1^^M{\expandafter\toks0\expandafter{\striprel@x #1}%
\edef\next{\the\toks0}\ifx\next\em@rk\let\next=\endgroup\else\ifx\next\empty%
\else\immediate\write\wfile{\the\toks0}\fi\let\next=\writ@line\fi\next\relax}}
\def\striprel@x#1{} \def\em@rk{\hbox{}}
\def\lref{\begingroup\obeylines\lr@f}
\def\lr@f#1#2{\gdef#1{\ref#1{#2}}\endgroup\unskip}

\def\addref#1{\immediate\write\rfile{\noexpand\item{}#1}} 
\def\startrefs#1{\immediate\openout\rfile=refs.tmp\refno=#1}
\def\xref{\expandafter\xr@f}\def\xr@f[#1]{#1}
\def\cite{\expandafter\cxr@f}\def\cxr@f[#1]{$^{#1)}$}
\def\refs#1{\count255=1$^{\r@fs #1{\hbox{}})}$}
\def\r@fs#1{\ifx\und@fined#1\message{reflabel \string#1 is undefined.}%
\nref#1{need to supply reference \string#1.}\fi%
\vphantom{\hphantom{#1}}\edef\next{#1}\ifx\next\em@rk\def\next{}%
\else\ifx\next#1\ifodd\count255\relax\xref#1\count255=0\fi%
\else#1\count255=1\fi\let\next=\r@fs\fi\next}
%

%
\newwrite\ffile\global\newcount\figno \global\figno=1
\def\fig{fig.~\the\figno\nfig}
\def\nfig#1{\xdef#1{fig.~\the\figno}%
\writedef{#1\leftbracket fig.\noexpand~\the\figno}%
\ifnum\figno=1\immediate\openout\ffile=figs.tmp\fi\chardef\wfile=\ffile%
\immediate\write\ffile{\noexpand\medskip\noexpand\item{Fig.\ \the\figno. }
\reflabeL{#1\hskip.55in}\pctsign}\global\advance\figno by1\findarg}
\def\xfig{\expandafter\xf@g}\def\xf@g fig.\penalty\@M\ {}
\def\figs#1{figs.~\f@gs #1{\hbox{}}}
\def\f@gs#1{\edef\next{#1}\ifx\next\em@rk\def\next{}\else
\ifx\next#1\xfig #1\else#1\fi\let\next=\f@gs\fi\next}
\newwrite\lfile
{\escapechar-1\xdef\pctsign{\string\%}\xdef\leftbracket{\string\{}
\xdef\rightbracket{\string\}}\xdef\numbersign{\string\#}}

\def\writestop{\def\writestoppt{\immediate\write\lfile{\string\pageno%
\the\pageno\string\startrefs\leftbracket\the\refno\rightbracket%
\string\def\string\secsym\leftbracket\secsym\rightbracket%
\string\secno\the\secno\string\meqno\the\meqno}\immediate\closeout\lfile}}
\def\writestoppt{}\def\writedef#1{}
\def\seclab#1{\xdef #1{\the\secno}\writedef{#1\leftbracket#1}\wrlabeL{#1=#1}}
\def\subseclab#1{\xdef #1{\secsym\the\subsecno}%
\writedef{#1\leftbracket#1}\wrlabeL{#1=#1}}
\newwrite\tfile \def\writetoca#1{}
\def\leaderfill{\leaders\hbox to 1em{\hss.\hss}\hfill}
\def\writetoc{\immediate\openout\tfile=toc.tmp
   \def\writetoca##1{{\edef\next{\write\tfile{\noindent ##1
   \string\leaderfill {\noexpand\number\pageno} \par}}\next}}}

%
\catcode`\@=12 
%
\edef\tfontsize{\ifx\answ\bigans scaled\magstep3\else scaled\magstep4\fi}
 \tfontsize  \tfontsize
 \tfontsize \font\titlei=cmmi10 \tfontsize
\font\titleis=cmmi7 \tfontsize \font\titleiss=cmmi5 \tfontsize
\font\titlesy=cmsy10 \tfontsize \font\titlesys=cmsy7 \tfontsize
\font\titlesyss=cmsy5 \tfontsize  \tfontsize
\skewchar\titlei='177 \skewchar\titleis='177 \skewchar\titleiss='177
\skewchar\titlesy='60 \skewchar\titlesys='60 \skewchar\titlesyss='60
 \ifx\answ\bigans\else scaled\magstep1\fi
%
\font\abssl=cmsl10 scaled 833
\font\absrm=cmr10 scaled 833 \font\absrms=cmr7 scaled  833
\font\absrmss=cmr5 scaled  833 \font\absi=cmmi10 scaled  833
\font\absis=cmmi7 scaled  833 \font\absiss=cmmi5 scaled  833
\font\abssy=cmsy10 scaled  833 \font\abssys=cmsy7 scaled  833
\font\abssyss=cmsy5 scaled  833 \font\absbf=cmbx10 scaled 833
\skewchar\absi='177 \skewchar\absis='177 \skewchar\absiss='177
\skewchar\abssy='60 \skewchar\abssys='60 \skewchar\abssyss='60
\def\abstractfont{\def\rm{\fam0\absrm}
\textfont0=\absrm \scriptfont0=\absrms \scriptscriptfont0=\absrmss
\textfont1=\absi \scriptfont1=\absis \scriptscriptfont1=\absiss
\textfont2=\abssy \scriptfont2=\abssys \scriptscriptfont2=\abssyss
\textfont\itfam=\absi \def\it{\fam\itfam\absi}\def\footnotefont{\tenpoint}%
\textfont\slfam=\abssl \def\sl{\fam\slfam\abssl}%
\textfont\bffam=\absbf \def\bf{\fam\bffam\absbf}\rm}
\font\ftsl=cmsl10 scaled 833
\font\ftrm=cmr10 scaled 833 \font\ftrms=cmr7 scaled  833
\font\ftrmss=cmr5 scaled  833 \font\fti=cmti10 scaled  833
\font\ftis=cmmi7 scaled  833 \font\ftiss=cmmi5 scaled  833
\font\fttt=cmtt10 scaled 833
\font\ftsy=cmsy10 scaled  833 \font\ftsys=cmsy7 scaled  833
\font\ftsyss=cmsy5 scaled  833 \font\ftbf=cmbx10 scaled 833
\skewchar\fti='177 \skewchar\ftis='177 \skewchar\ftiss='177
\skewchar\ftsy='60 \skewchar\ftsys='60 \skewchar\ftsyss='60
\def\footnotefont{\def\rm{\fam0\ftrm}
\textfont0=\ftrm \scriptfont0=\ftrms \scriptscriptfont0=\ftrmss
\textfont1=\fti \scriptfont1=\ftis \scriptscriptfont1=\ftiss
\textfont2=\ftsy \scriptfont2=\ftsys \scriptscriptfont2=\ftsyss
\textfont\itfam=\fti \def\it{\fam\itfam\fti}\def\footnotefont{\tenpoint}%
\textfont\slfam=\ftsl \def\sl{\fam\slfam\ftsl}
\def\tt{\fam\ttfam\fttt}%
\textfont\bffam=\ftbf \def\bf{\fam\bffam\ftbf}\rm}
%
\def\tenpoint{\def\rm{\fam0\tenrm}
\textfont0=\tenrm \scriptfont0=\sevenrm \scriptscriptfont0=\fiverm
\textfont1=\teni  \scriptfont1=\seveni  \scriptscriptfont1=\fivei
\textfont2=\tensy \scriptfont2=\sevensy \scriptscriptfont2=\fivesy
\textfont\itfam=\tenit \def\it{\fam\itfam\tenit}\def\footnotefont{\ninepoint}%
\textfont\bffam=\tenbf \def\bf{\fam\bffam\tenbf}\def\sl{\fam\slfam\tensl}\rm}
\font\ninerm=cmr9 \font\sixrm=cmr6 \font\ninei=cmmi9 \font\sixi=cmmi6
\font\ninesy=cmsy9 \font\sixsy=cmsy6 \font\ninebf=cmbx9
\font\nineit=cmti9 \font\ninesl=cmsl9 \skewchar\ninei='177
\skewchar\sixi='177 \skewchar\ninesy='60 \skewchar\sixsy='60
\def\ninepoint{\def\rm{\fam0\ninerm}
\textfont0=\ninerm \scriptfont0=\sixrm \scriptscriptfont0=\fiverm
\textfont1=\ninei \scriptfont1=\sixi \scriptscriptfont1=\fivei
\textfont2=\ninesy \scriptfont2=\sixsy \scriptscriptfont2=\fivesy
\textfont\itfam=\ninei \def\it{\fam\itfam\nineit}\def\sl{\fam\slfam\ninesl}%
\textfont\bffam=\ninebf \def\bf{\fam\bffam\ninebf}\rm}
%
%

\input epsf
\tolerance=10000
\hfuzz=5pt
\baselineskip 12truept plus 0.5truept minus 0.5truept
\pageno=0\nopagenumbers\tolerance=10000\hfuzz=5pt
\line{\hfill CERN-TH/95-305}
\line{\tt\hfill hep-ph/9511345}
\vskip 24pt
\centerline{\bf POLARIZED STRUCTURE FUNCTIONS:}
\centerline{\bf A THEORETICAL UPDATE}
\vskip 36pt
\centerline{Stefano~Forte\footnote*{\footnotefont\baselineskip6pt
On leave
 from INFN, Sezione di Torino, via P.~Giuria 1,
I-10125 Turin, Italy (address after December 1, 1995).}}
\vskip 12pt
\centerline{\it Theory Division, CERN,}
\centerline{\it CH-1211 Gen\`eve 23, Switzerland.}
\vskip 48pt
{\narrower\baselineskip 10pt
\centerline{\bf Abstract}
\medskip
We review recent developments in the theory and phenomenology of polarized
structure functions. We summarize recent experimental data on the proton
and deuteron structure function $g_1$, and their impact on the understanding
of polarized sum rules. Specifically,
we discuss how accurate  measurements of the singlet and nonsinglet first
moment of $g_1$ test perturbative and nonperturbative QCD, and critically
examine
the way these measurements are arrived at.
We then discuss how the extraction of structure
functions from the data can be improved by means of a theoretical
analysis of their $x$ and $Q^2$ dependence, and how, conversely, experimental
information on this dependence can be used to pin down the polarized parton
content of the nucleon.
\smallskip}
\vskip 50pt
\centerline{Invited talk at the VI Blois Workshop}
\centerline{\it Frontiers in Strong Interactions}
\centerline{Blois, France, June 1995}
\medskip
\centerline{\it to be published in the proceedings}
\vfill
\line{CERN-TH/95-305\hfill}
\line{November 1995\hfill}
\eject
\footline={\hss\tenrm\folio\hss}
\def\neath#1#2{\mathrel{\mathop{#1}\limits_{#2}}}
\def\lsim{\mathrel{\rlap{\lower4pt\hbox{\hskip1pt$\sim$}}
    \raise1pt\hbox{$<$}}}         
\def\gsim{\mathrel{\rlap{\lower4pt\hbox{\hskip1pt$\sim$}}
    \raise1pt\hbox{$>$}}}         
\def\epm#1#2{\hbox{${+#1}\atop {-#2}$}}
\def\MS{\hbox{$\overline{\rm MS}$}}
\def\PR{{\it Phys.~Rev.~}}
\def\PRL{{\it Phys.~Rev.~Lett.~}}
\def\NP{{\it Nucl.~Phys.~}}
\def\NPBPS{{\it Nucl.~Phys.~B (Proc.~Suppl.)~}}
\def\PL{{\it Phys.~Lett.~}}

\def\ZP{{\it Z.~Phys.~}}

\def\vol#1{{\bf #1}}
\def\vyp#1#2#3{\vol{#1} (#2) #3}

\def\tr{\,{\hbox{tr}}\,}

\def\frac#1#2{{{#1}\over {#2}}}
\def\half{\hbox{${1\over 2}$}}

\def\smallfrac#1#2{\hbox{${{#1}\over {#2}}$}}
\def\tr{{\rm tr}}

\baselineskip 15truept plus 0.5truept minus 0.5truept
\nref\smcp{SMC Collaboration,
D.~Adams et al., \PL\vyp{B329}{1994}{399}.}
\nref\slacp{E143
Collaboration, K.~Abe et al., \PRL\vyp{74}{1995}{346}.}
\nref\smcd{ SMC Collaboration,
D.~Adams et al., \PL\vyp{B357}{1995}{248}.}
\nref\slacd{E143
Collaboration, K.~Abe et al., \PRL\vyp{75}{1995}{25}.}
\newsec{The state of the art}
The state of the knowledge of polarized structure functions
measured in deep-inelastic scattering has progressed
during the
last couple of years from the level of testing naive
parton model ideas to the point of probing QCD at leading
order (LO) and next-to-leading
order (NLO). On the one hand,
experiments at CERN and SLAC\refs{\smcp-\slacd}
have accurately measured the structure function $g_1$, which
determines the leading twist contribution to the longitudinal
polarization asymmetry,
for proton and deuteron targets,
 over a reasonably wide range in $x$ and at a pair of values of $Q^2$.
On the other hand, the perturbative
behavior of $g_1$
in the $(x,Q^2)$ plane  can
now be quantitatively studied at NLO\ref\BFRb{R.~D.~Ball, S.~Forte and
G.~Ridolfi, preprint CERN-TH/95-266, {\tt hep-ph/9510449}.},
thanks to the recent
determination
of the full set of two-loop polarized anomalous
dimensions\ref\merner{R.~Mertig and W.~L.~van~Neerven, Leiden preprint
INLO-PUB-6/95; Erratum, private communication (November 1995).}.
This means that polarized structure function data can now be used to
test perturbative QCD and obtain detailed information on the
structure of polarized nucleons, while, conversely
the data can be understood and analyzed using accurate NLO methods.

Here we will review these recent developments, emphasizing the impact of the
new data on the theoretical understanding of polarized structure functions.
As it often happens, quantities that  are easier to understand theoretically
are the hardest to measure, and vice versa. We will take a theorist's
viewpoint: we will start from the experimental results which have
the most direct theoretical interpretation, and discuss them
without, at first,
questioning the procedure through which they have been obtained;
we will then proceed to quantities whose interpretation is more subtle,
and eventually discuss the procedure which is used to extract information
from the data.

In Sect.~2 we will review the determinations of the first moment of
$g_1$ (as given by the experimental collaborations) and discuss
in particular the behavior of the isotriplet first moment for
which there is an absolute QCD prediction (the Bjorken sum rule).
In Sect.~3 we will turn to the singlet component of the first
moment, discuss its extraction from
the data and its theoretical interpretation, which is complicated
by the
presence in this channel of the axial anomaly. In Sect.~4 we will
then turn to the full $x,Q^2$ dependence of the
structure function $g_1$; we will discuss how it is described
in perturbative QCD, and how it may be used
to obtain  information on the general
structure of polarized parton distributions. Finally, we will
come full circle, and  show
how this information can be used to improve the precision of the extraction
from the data of quantities of simple and direct theoretical
relevance, such as the moments of parton distributions.

In this review we will concentrate on the physics of
$g_1$\ref\revsf{For a general
introduction
to  the phenomenology of $g_1$, see for instance
S.~Forte,
{\tt hep-ph/9409416}, in ``Radiative
Corrections: Status and Outlook'', B.~F.~L.~Ward, ed. (World Scientific,
Singapore, 1995) and references
therein. We will follow the notation and conventions
of this paper.},
the discussion of other polarized structure functions being outside
the scope of our treatment. It is however worth mentioning that
first measurements of $g_2$, the other structure function which
contributes to the deep-inelastic polarized cross section
have been performed recently\ref\gtwo{SMC
Collaboration,
D.~Adams et al., {\it Phys}. {Lett}. {\bf B336} (1994) 125;
E143
Collaboration, K.~Abe et al., preprint
SLAC-PUB-95-6982.}. While the theoretical interpretation of this
structure function is worth studying for its own
sake\ref\alnr{See e.g. R.~L.~Jaffe,
{\it Comm. Nucl. Part.
Phys.}
{\bf 14} (1990) 239; G.~Altarelli,
B.~Lampe, P.~Nason and G.~Ridolfi, {\it Phys}. {\it Lett}. {\bf B334} (1994)
187 and references therein.},
its contribution to  the longitudinally
polarized cross-section, on which we will concentrate,
vanishes asymptotically (as $1\over Q^2$). In this contex,  it is thus
  important mostly as a background:
the recent experiments find that $g_2$
is rather small, and essentially compatible with zero
within errors, thus supporting the view that $g_1$ may be accurately
determined by simply neglecting $g_2$.
\bigskip
\newsec{The nonsinglet first moment and the Bjorken sum rule}
\medskip
The main outcome of the recent precise experiments\refs{\smcp-\slacd}
is a set of determinations of the first moment
\eqn\fmom{\Gamma_1(Q^2)=\int_0^1\!dx\,g_1(x,Q^2)}
for proton and deuteron targets, displayed in
table~1.\footnote*{\footnotefont\baselineskip6 pt
There exists also  a direct determination of the neutron
structure function and its first moment, obtained from scattering on a {}$^3$He
target\ref\neut{E142 Collaboration, P.~L.~Anthony et al., {\it
Phys}. {\it Rev}. {\it Lett}. {\bf 71} (1993) 959.}.
We will not include this result in
our discussion, because of its large statistical and systematic
 uncertainties --- partly due to the problems related to scattering on a
nuclear
target\cite\revsf  ---
and also because a recent reanalysis of the same raw data
has lead to a rather different determination of the
corresponding structure function\ref\reneu{Y.~Roblin, Ph.D. Thesis,
University of
Clermont--Ferrand (1995).}.}
We will discuss later the theoretical input that goes into these
determinations. If we take them at face value, however, they
provide us directly with information that admits  a simple
theoretical interpretation, because $\Gamma_1$ measures the nucleon
matrix element of the axial current:\eqnn\axme\eqnn\gamdec
$$
\eqalignno{M^t a_i^ts^\mu&\equiv \langle t; p, s |\bar \psi_i \gamma_\mu
\gamma_5 \psi_i
| t; p, s \rangle,&\axme\cr
\Gamma^{t}_1(Q^2)&=
{1\over 2} \sum_{i=1}^{n_f} e^2_i C_i(Q^2)
a_i^t={1\over 2}\left[\langle e^2\rangle C_{\rm S}(Q^2)a^t_0(Q^2)+
C_{\rm NS}(Q^2)a^t_{\rm NS}(Q^2)
\right],&\gamdec\cr}
$$
where $t$ indicates the target hadron (proton or deuteron, in our case)
with mass $M^t$, momentum $p^\mu$ and spin $s^\mu$, and
the sum runs over all activated quark flavors;
in the sequel when dropping the label $t$ we will tacitly
refer to a proton target.
In the last step we have
introduced  the average quark charge $\langle e^2\rangle={1\over n_f}
\sum_{i=1}^n e^2_i$,
and the singlet and nonsinglet axial charges
\eqnn\asi\eqnn\ansi
$$\eqalignno{
a_0 &=\sum_{i=1}^{n_f} a_i&\asi\cr
a^{\rm NS}&=\sum_{i=1}^{n_f}
\left(e_i^2-\langle e^2\rangle\right) a_i,&\ansi\cr}$$
exploiting the fact that
the Wilson coefficient functions $C(Q^2)$ only distinguish between singlet
and nonsinglet currents.
\topinsert\hfil
\vbox{\tabskip=0pt \offinterlineskip
      \def\tablerule{\noalign{\hrule}}
      \halign to 350pt{\strut#&\vrule#\tabskip=1em plus2em
                   &\hfil#\hfil&\vrule#
                   &#\hfil&\vrule#
                   &#\hfil&\vrule#
                   &#\hfil&\vrule#
                   &\hfil#&\vrule#\tabskip=0pt\cr\tablerule
      &&\omit\hidewidth Ref.\hidewidth
      &&\omit\hidewidth Target\hidewidth
      &&\omit\hidewidth $\langle Q^2\rangle $ \hidewidth
      &&\omit\hidewidth $\Gamma_1$\hidewidth
             &&\omit\hidewidth $a_0$\hidewidth&\cr\tablerule
    && \xref\smcp&&p && 10 && $0.136\pm0.011\pm0.011$ &&
              $0.22\pm0.14$ &\cr\tablerule
   && \xref\slacp &&p&& 3 && $0.129\pm0.004\pm0.009$ &&
              $0.29\pm0.10$ &\cr\tablerule
   && \xref\smcd &&d&& 10 && $0.034\pm0.009\pm0.006$ &&
              $0.20\pm 0.11$ &\cr\tablerule
   && \xref\slacd &&d&& 3 && $0.042\pm0.003\pm 0.004$ &&
            $0.30\pm 0.06$ &\cr\tablerule}}
\hfil\bigskip\noindent{\abstractfont\baselineskip6pt\narrower
Table 1: Summary of recent experimental determinations of $\Gamma_1$
eq.~\fmom\
and $a_0$ eq.~\asi\ (as presented by the experimental collaborations).
The first error
on $\Gamma_1$ is statistical and the second systematic.
All results hold at the  average scale $\langle Q^2\rangle$
(in GeV$^2$)
of the respective experiments. \smallskip}
\medskip
\endinsert

The nonsinglet axial charge is particularly simple because the corresponding
current is conserved (neglecting quark masses) and can therefore
be decomposed into
a sum of scale--independent contributions:
\eqn\nsingdec{\eqalign{
a_{\rm NS}(Q^2)&
=\sum_{k=2}^{n_f}\left[\left(\langle e^2\rangle_{k-1}
-
\langle e^2\rangle_{k}\right)\Theta(Q^2-Q^2_k)\left(\left[\sum_{i=1}^{k-1}
a_i\right]-
(k-1) a_k\right)\right]\cr
&={1\over 6} \left(a_u-a_d\right)+
{1\over 18} \left(a_u+a_d-2 a_s\right)\cr&\qquad\quad
-{1\over 18}\Theta(Q^2-Q^2_c)
\left( a_u+a_d+a_s-3a_c\right)+\dots\cr}}
where $\langle e^2\rangle_{k}$ is the average charge computed with $n_f=k$
flavors, $ Q^2_c$ is the threshold for the $k$-th flavor, and
all the scale dependence comes from the function
$\Theta(Q^2-Q^2_k)$, which vanishes below
threshold and is equal to 1 above threshold
(having a nucleon target in mind in the last step
we explicitly displayed thresholds for the charm and
heavier quarks).

The simplest nonsinglet
scale-independent currents are then those related to light quarks,
i.e. the isotriplet
axial charge $g_A$ and the SU(3) octet charge $a_8$
\eqn\tripoc
{g_A=a_u-a_d;\qquad a_8=a_u+a_d-2a_s.}
Using isospin symmetry, knowledge of $\Gamma_1$ for proton and neutron
targets
determines the isotriplet
combination:
\eqn\bjsr
{\eqalign{\Gamma_1^{I=1}\equiv\Gamma_1^p-\Gamma_1^n=&C_{\rm NS}(Q^2)
{1\over 6} \left(a_u^p-a_d^p\right)\cr=&
\bigg[1-\left({\alpha_s\over\pi}\right)
-\left(\smallfrac{55}{12}-\smallfrac{1}{3}n_f\right)
\left({\alpha_s\over\pi}\right)^2\cr&\qquad-
\left(41.4399-7.6072 n_f+
\smallfrac{115}{648}n_f^2\right)
\left({\alpha_s\over\pi}\right)^3+O(\alpha_s^4)\bigg]{1\over 6} g_A.\cr}}
 The only assumption that goes into the derivation of eq.~\bjsr\ is that
of exact isospin symmetry, which implies $a_u^p=a_d^n$, $a_d^p=a_u^n$
and $a_h^p=a_h^n$, where $h$ indicates strange or heavy quark flavors.
Furthermore, using isospin algebra again, $g_A$ in eq.~\tripoc\ can be related
to the matrix element of the axial current which induces beta decay,
and thus to the beta decay constant of the same name,
which is known rather accurately: $g_A=1.2573\pm 0.0028$.
The relation eq.~\bjsr\ between the beta decay constant and
the isotriplet first moment of $g_1$ is known as the Bjorken sum
rule\ref\bj{J.~D.~Bjorken, {\it Phys}. {\it Rev}. {\bf 148} (1966) 1467}.
In practice, the neutron structure function can be extracted from the deuteron
one by using additivity of the deep-inelastic cross sections after
correcting for nuclear effects\ref\frastri{L.~L.~Frankfurt and M.~Strikman,
{\it Nucl}. {\it Phys}. {\bf A405} (1983) 557.}:
\eqn\deunuc{\Gamma_1^d=(1-1.5\omega_D){\Gamma_1^p+\Gamma_1^n\over2},}
where $\omega_D\approx0.05$ is the probability that the deuteron is
in a D-wave state.

It thus looks as though the Bjorken sum rule \bjsr\ is an ideal place to
test QCD in a clean and simple way: in principle, with a pair of determinations
of $\Gamma_1^{I=1}$ at different scales both its normalization
and its scale dependence may be measured, and thus isospin symmetry (which
determines the value of $g_A$) and perturbative QCD [which
determines $C_{\rm NS}(Q^2)$]  may be simultaneously tested. Now,
isospin symmetry violation
in QCD is
 expected to be suppressed by powers of  the current quark mass
on a typical strong interaction scale;  in particular
in the unpolarized quark distributions it is at most of a few
per cent~\ref\iso{S.~Forte, {\it Phys}. {\it Rev}.
{\bf D47} (1993) 1842.}; the observation of larger isospin violation
 in this channel would presumably be the sign of nonperturbative
physics and rather interesting per se~\ref\sfbj{S.~Forte, {\it Phys}.
{\it Lett}.
{\bf B309} (1993) 174.}.

On the other hand,
$C_{\rm NS}(Q^2)$ is known up to
order $\alpha_s^3$, i.e., at the four--loop order, thereby allowing very
precise tests of perturbation theory.  In fact, at this order it might already
be possible to see\ref\pade{J.~Ellis et al., preprint CERN-TH/95-155,
{\tt hep-ph/9509312}.} manifestations of the fact
that the perturbative expansion in powers of $\alpha_s$ of QCD observables
diverges\ref\braun{See e.g. V.~Braun,
{\tt hep-ph/9505317} and references therein for a simple phenomenological
discussion.}. A detailed treatment of this issue is beyond
the scope of the present paper; its main implication for our
purposes is that because of its divergent nature the perturbative
series is ambiguous, in that different resummations of the series may differ
by terms which are proportional to powers of $\Lambda^2_{\rm QCD}\over Q^2$.
Because the values of physical observables are unambiguous,
these ambiguities must cancel against corresponding ones in
other power-suppressed contributions, i.e. higher twist
contributions.

Indeed, in general eq.~\bjsr\ should read
\eqn\bjht
{\Gamma_1^{I=1}={1\over 6}\left[ C_{NS}(Q^2)g_A+{C_{\rm HT}\over
Q^2}+O\left({1\over Q^4}\right)\right].}
The coefficient $C_{\rm HT}$ can  be computed  once
a prescription for the treatment of the perturbative expansion is specified;
the result for $\Gamma_1$
is then unambiguous\ref\fran{F.~David, {\it Nucl}. {\it Phys}.
{\bf B263}
 (1986) 637;  A.~H.~Mueller,  {\it Nucl}. {\it Phys}. {\bf B250}
 (1985) 327.}. This program has never been carried through explicitly
(there exists a proposal\ref\ji{X.~Ji, {\it Nucl}. {\it Phys}. {\bf B448}
(1995)
51.}  based on an explicit cutoff scheme which however
would
require a determination of $C_{\rm HT}$ on the lattice). It is
nevertheless possible to estimate the size of $C_{\rm HT}$
phenomenologically, for instance
using QCD sum rules; this  leads to\ref\htsr{For a
comprehensive summary of the determinations of
higher twist corrections to the Bjorken
sum rule see
L.~Mankiewicz,
E.~Stein and A.~Sch\"afer, {\tt hep-ph/9510418}.}
$C_{\rm HT}\approx -0.1$ with an error of order 50\% or more.
Such a determination,
however, is  only meaningful if  the ambiguity
is substantially smaller than the value of $C_{\rm HT}$.
Even though a qualitative argument\cite\braun\ suggests that this is the case,
an
estimate\cite\pade\ based
on an analysis of the known terms in the expansion of $C_{\rm NS}$
with the method of Pad\'e  approximants indicates instead that
the ambiguity is of the same size as $C_{\rm HT}$.

Be that as it may, these estimates  give an
indication of the level of accuracy at which
perturbative QCD can be tested by
comparing the Bjorken sum rule with the data. This comparison is shown in
fig.~1,
where $\Gamma_1^{I=1}$ is displayed as a function of $Q^2$ at various
perturbative orders, with and without typical higher twist corrections,
for different values of the strong coupling, and compared with  an
experimental value
obtained\cite\pade\ by averaging all the data of table~1
(plus some less precise earlier data) evolved to a common scale
(we will discuss this evolution in the next section):
\eqn\bjelkar{\Gamma_1^{I=1}(\hbox{3 GeV}^2)=0.164\pm0.011.}
The comparison
is striking: the result appears to be very sensitive to perturbative
corrections --- it would disagree with the theoretical prediction if this
were computed at LO rather than NNLO ---
while being reasonably insensitive to higher twists
and associated ambiguities. In fact,  the value
of $\alpha_s$ appears to be tested here
to an accuracy comparable with the world
average of all other determinations\ref\elkaras{J.~Ellis and M.~Karliner,
{\it Phys. Lett.}
{\bf B341} (1995) 397.}. In other words, if, rather than
assuming isospin and using eq.~\bjsr\ to measure $\alpha_s$, we
assign the value of $\alpha_s$ and test isospin, then
 the
uncertainty
due to the error on the value
of $\alpha_s$ is already larger than the experimental
uncertainty on the value of $\Gamma_1^{I=1}$ quoted in eq.~\bjelkar.
\topinsert
\epsfysize=16truecm\vskip-3.8truecm
\hfil\epsfbox{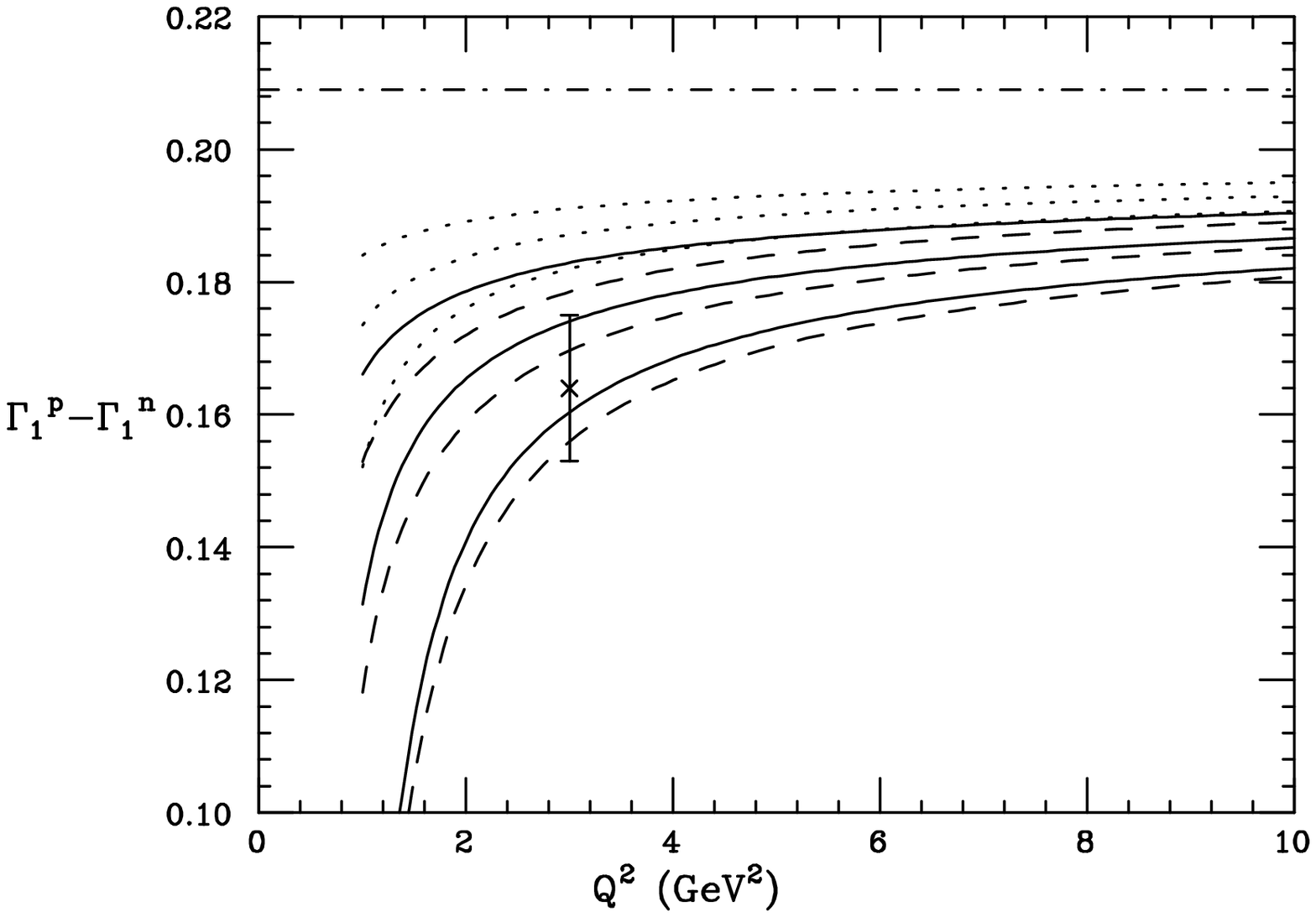}\hfil\vskip-4.2truecm
\bigskip\noindent{\abstractfont\baselineskip6pt\narrower
Figure 1: Scale dependence of $\Gamma_1^{I=1}$ eq.~\bjht\
[adapted from ref.~\ref\anr{G.~Altarelli, P.~Nason and
G.~Ridolfi, {\it Phys}. {\it Lett}. {\bf B320} (1994) 152.}]. The dashed curve
corresponds to eq.~\bjsr\ and the solid curve to eq.~\bjht\ with
$C_{\rm HT}=-0.1$; the dotted curve displays the $O(\alpha_s)$ contribution
to eq.~\bjsr, and the dotdashed is the asymptotic ($\alpha_s=0$) value.
The three sets of curves correspond to the values
$\alpha_s(M_z)=0.118\pm0.007$.
The data point is as in eq.~\bjelkar.
\smallskip}
\medskip
\endinsert

This determination
of $\Gamma_1^{I=1}$, however, while being a theorist's dream,
is a phenomenologist's nightmare, and should be taken with extreme
care. Firstly, the inclusion of nuclear effects in eq.~\deunuc\ as a simple
multiplicative correction factor may be an oversimplification:
in a more detailed treatment of the deuteron
wave function the correction is actually
 $x$ dependent, and its effect on  the first moment
 depends on the shape of the structure
function $g_1(x)$; the effect has been argued to  be non--negligible
at the level of present-day accuracy, especially in the large--$x$
region\ref\thom{W.~Melnitchouk, G.~Piller and A.~W.~Thomas,
 \PL {\bf B346} (1995) 165.}. Other nuclear effects,
such
as shadowing, which has a substantial effect\nref\bad{B.~Bade\l ek and
J.~Kwieci\`nski, {\it Phys}. {\it Rev}. {\bf D80} (1994) 4.}\refs{\iso,\bad}
on the unpolarized
counterpart of $\Gamma_1^{I=1}$ (the isotriplet first moment of $F_2$),
have been pointed out in this context\ref\shad{N.~N.~Nikolaev and
V.~I.~Zakharov, {\it Phys}. {\it Lett}.
{\bf B55} (1975) 397;  S.~J.~Brodsky and H.~J.~Lu, {\it Phys}. {\it Rev}.
{\it Lett}. {\bf 64} (1990) 1342.}, but not studied systematically.

Furthermore,
it is clear that the reason why this test of perturbative QCD appears
to be so sensitive is that the measurement is performed at
a low scale, which exponentially magnifies the sensitivity to perturbative
evolution.  At such a low scale, however, other sorts of higher twist
corrections, besides those discussed above, could be relevant,
in particular those related to the nucleon mass: even assuming
that corrections due to transverse polarization (i.e. proportional to $g_2$)
are negligible, there are still kinematic higher twist corrections
such as target mass corrections\ref\tmc{O.~Nachtmann, {\it Nucl}. {\it Phys}.
{\bf B63} (1973) 237;  S.~Wandzura, {\it Nucl}. {\it Phys}.
{\bf B122} (1977) 412;  S.~Matsuda and T.~Uematsu, {\it Nucl}. {\it Phys}.
{\bf B168} (1980) 181.}. Their effect on the first moment of $\Gamma_1^{I=1}$
is\ref\bjtmc{H.~Kawamura and
T.~Uematsu, {\it Phys}. {\it Lett}. {\bf B343} (1995) 346.} comparable to the
uncertainty in eq.~\bjelkar\ (which does not include it)
at that scale. Moreover,
such effects have never been studied systematically and
could affect the extraction of all the first moments of table~1 from
the data: even determinations with large $\langle Q^2\rangle$
have several data points at low $Q^2$.
Finally, the result \bjelkar\ is obtained by evolving the full first moment
$\Gamma_1$, which in turn is obtained from data taken at various values of
$Q^2$. The corresponding uncertainty, related to
perturbative evolution, could also be rather large at low scales:
this will be the main subject of sect.~4.

In conclusion, the fact that $\Gamma_1^{I=1}$ is only measured
at relatively low $Q^2$ allows testing of the Bjorken sum rule
only to about 10\% accuracy, due to the large sensitivity
to the value of $\alpha_s$. This could be turned around: assuming
the validity of the sum rule (i.e. of exact isospin) the value of $\alpha_s$
can in principle be measured rather
accurately. There is however a trade off,
in that the determination of $\Gamma_1^{I=1}$ at a low scale is
affected by substantial systematic uncertainties, and the errors
in table~1 may turn out to be over--optimistic, as we will see
explicitly in Sect.~4.
\bigskip
\newsec{The singlet axial charge and the  anomaly}
\medskip
The singlet component of $\Gamma_1$ can be extracted from the data
in several  distinct ways:
by taking suitable linear combinations of different experiments, or
by using SU(3) symmetry, or by exploiting the fact that the
scale dependence of the
singlet first moment differs from that of the nonsinglet.
This scale dependence must also be taken into account when comparing
different experiments, and turns out to pose some interesting theoretical
questions. We will discuss the phenomenological and theoretical aspects
of these issues in turn.
\medskip
\subsec{The first moment and its scale dependence}
\smallskip
The simplest way of getting a value for the singlet first moment
and the associated charge $a_0(Q^2)$ [eq.~\asi] is to
subtract the nonsinglet contribution off each experimental
determination of $\Gamma_1$. The axial charge is then found  from
eq.~\gamdec\ by dividing out the singlet coefficient function,
which has been computed\ref\lar{S.~A.~Larin, {\it Phys}. {\it Lett}.
{\bf B334} (1994) 192.} to NNLO:
\eqn\singcf{C_{\rm S}(Q^2)=
\left[1-\left({\alpha_s\over\pi}\right)
-\left(\smallfrac{55}{12}-1.16248 n_f\right)
\left({\alpha_s\over\pi}\right)^2+O(\alpha_s^3)\right].}
The values
listed in table~1 were obtained in this way.

Neglecting all contributions from
heavy quarks (we will come back to this assumption later)
the nonsinglet is  given by
\eqn\nsing
{\Gamma_1^{\rm nonsing.}={1\over 6}C_{\rm NS}(Q^2) \left(g_A+{1\over 3}
a_8\right).}
Just as $g_A$ can be obtained from the nucleon beta decay constants
using isospin, $a_8$ can
be obtained from hyperon beta decay constants using SU(3)
symmetry, with the result $a_8=0.579\pm0.025$.
This value is arrived at\ref\cloro{F.~E.~Close and R.~G.~Roberts,
{\it Phys}. {\it Lett}. {\bf B336} (1994) 257.} by a
best fit based on the assumption of exact
SU(3) symmetry; it could therefore by significantly affected by SU(3)
symmetry breaking. This of course can only be introduced in a model-dependent
way. One possibility is to introduce a parametrization of the
hyperon beta decay constants which includes a
term proportional to SU(3) breaking in the octet mass spectrum.
A specific parametrization~\ref\ehrn{B.~Ehrnsperger and
A.~Sch\"afer, {\it Phys}. {\it Lett}. {\bf B348} (1995) 619.}
then leads
to the value $a_8=0.40\pm0.19$. An alternative option is to allow
for current mixing effects, i.e. let the singlet current have a nonvanishing
SU(3) nonsinglet matrix element, whose size can then be estimated within
a model for the quark content of the octet baryons\ref\lip{H.~J.~Lipkin,
{\it Phys}. {\it Lett}. {\bf B337} (1994) 157;  J.~Lichtenstadt and
H.~J.~Lipkin,
{\it Phys}. {\it Lett}. {\bf B353} (1995) 119.}, leading to  $a_8\approx0.4$
(and an uncertainty not smaller than the above)
when SU(3) breaking is maximal.

Even though the precise value of $a_8$ is thus affected by a large
uncertainty,
this has a rather small effect on the extraction of the singlet component,
because its numerical coefficient is one third of that of
the triplet and one fourth of the singlet. In fact, using these
values of $g_A$ and $a_8$ in the decomposition \gamdec\ of
the first moment into its singlet and nonsinglet components, it is easy to
see that about 90\% of the proton first moment comes from the isotriplet
component, and of the remaining 10\% about two-thirds are singlet, the
rest being octet. For the deuteron the isotriplet contribution
vanishes and the singlet and octet are partitioned as in the proton.
This trivial numerology has some important consequences
for the extraction of $a_0$ from the data: a)
if the singlet is extracted from the proton data
the uncertainty on
$a_0$ will be about ten times larger than that on
$\Gamma_1$; b) an uncertainty of  about 50\%
in the knowledge of $a_8$
has the same effect as an uncertainty of a few per cent on the
knowledge of the proton's $\Gamma_1$, i.e. it
generates an uncertainty of about 10\% in the proton or deuteron $a_0$;
c) the deuteron is in principle a better
probe of the singlet component.
This is reflected by the values of $a_0$
in table~1, derived
neglecting SU(3) breaking.

In order to meaningfully compare values of $a_0$ obtained in different
experiments, they must be evolved to the same scale.
Indeed, the singlet axial current is not conserved because
of the axial anomaly, as we will discuss more extensively in
the next section, so its matrix elements
can acquire an anomalous dimension, which
actually vanishes at LO, and has been
determined up to NNLO\cite\lar:
\eqn\azevol
{a_0(Q^2)= \left[1+\smallfrac{6n_f}{33-2n_f}\left({\alpha_s\over\pi}\right)+
\smallfrac{3087 n_f+ 138 n_f^2+4 n_f^3}{12(33-2n_f)^2}
\left({\alpha_s\over\pi}\right)^2+O(\alpha_s^3)
\right]a_0(\infty).}
The full scale dependence of the singlet contribution to $\Gamma_1$
[eq.~\gamdec]
is obtained by combining this with the singlet coefficient function
eq.~\singcf.

It is then possible to determine $a_0$ from a global fit to the data
of table~1  (rather
than by simply averaging the determinations coming from each data set):
if $g_A$ and $a_8$ are fixed using SU(3) symmetry as discussed
above, $a_0(\infty)$ is then the only free parameter. A global fit
(including also  as older data which however
carry little weight due to their poor accuracy)
with $\alpha_s(M_z)=0.118\pm0.007$ leads to\ref\alrid{G.~Altarelli
and G.~Ridolfi,
\NPBPS\vyp{39B}{1995}{106} and private communication.}
\eqn\guisig
{a_0(\infty)=0.29\pm0.04,}
where the error is almost entirely statistical, because of the
magnification of the statistical error on $\Gamma_1$
when $a_0$ is extracted.

Given the
availability of independent determinations of $\Gamma_1$
at different scales and for different targets, it
is actually possible to relax the assumptions that went into the
determination
eq.~\guisig. First,
it is possible to relax the assumption of SU(2) symmetry, and
extract simultaneously $a_0$ and $g_A$\cite\alrid:
\eqn\guisigga{a_0(\infty)=0.30\pm0.04;\quad g_A=1.12\pm0.10\epm{0.10}{0.04}.}
The error on $g_A$ has been decomposed in statistical and systematic;
the latter
is entirely due to the uncertainty on $\alpha_s$, the uncertainty
on $a_8$ being essentially negligible.  This shows, consistent
with the discussion in the previous section and fig.~1, that the Bjorken
sum rule can be tested at best to about 10\% accuracy, due to
the uncertainty on the value of $\alpha_s$.
 Finally, it should  in
principle be possible to determine simultaneously $g_A$, $a_8$ and
$a_0$ from the data: the proton--deuteron comparison fixes separately
the isotriplet (i.e. $g_A$) while the contributions of $a_0$ and  $a_8$
to the isosinglet could then be separated by their different scale
dependence. This is however not yet feasible in practice because
this scale dependence is only rather slight, while
the sensitivity of the value of $\Gamma_1$ to $a_8$ is weak; the latter
fact, however, implies that even if we were to assume a 30\% variation
of $a_8$ due to SU(3) violation the value of $a_0$ \guisig-\guisigga\
would hardly be affected.

We now turn to polarized
heavy quark contributions, which can be generated dynamically
by assuming that the corresponding current matrix elements vanish
on threshold. This then determines the various scale-independent
nonsinglet contribution
to eq.~\nsingdec\ in terms of the singlet at each threshold: for instance,
the charm contribution is
\eqn\ccont{
a_u+a_d+a_s-3a_c=[a_u+a_d+a_s](Q^2_c)=a_0(Q_c^2).}
Thus, because of the small numerical value of $a_0$, the error
made neglecting heavy quark contributions to the nucleon $\Gamma_1$
is indeed rather small.

We can finally compare all the results of table~1 to one another and to the
Bjorken sum rule prediction of eq.~\bjsr\ by evolving to a common scale.
The result, shown in fig.~2, demonstrates the mutual consistency of various
experiments and the agreement with the prediction of exact isospin
as tested by the Bjorken sum rule.
\topinsert
\epsfysize=16truecm\vskip-3.8truecm
\hfil\epsfbox{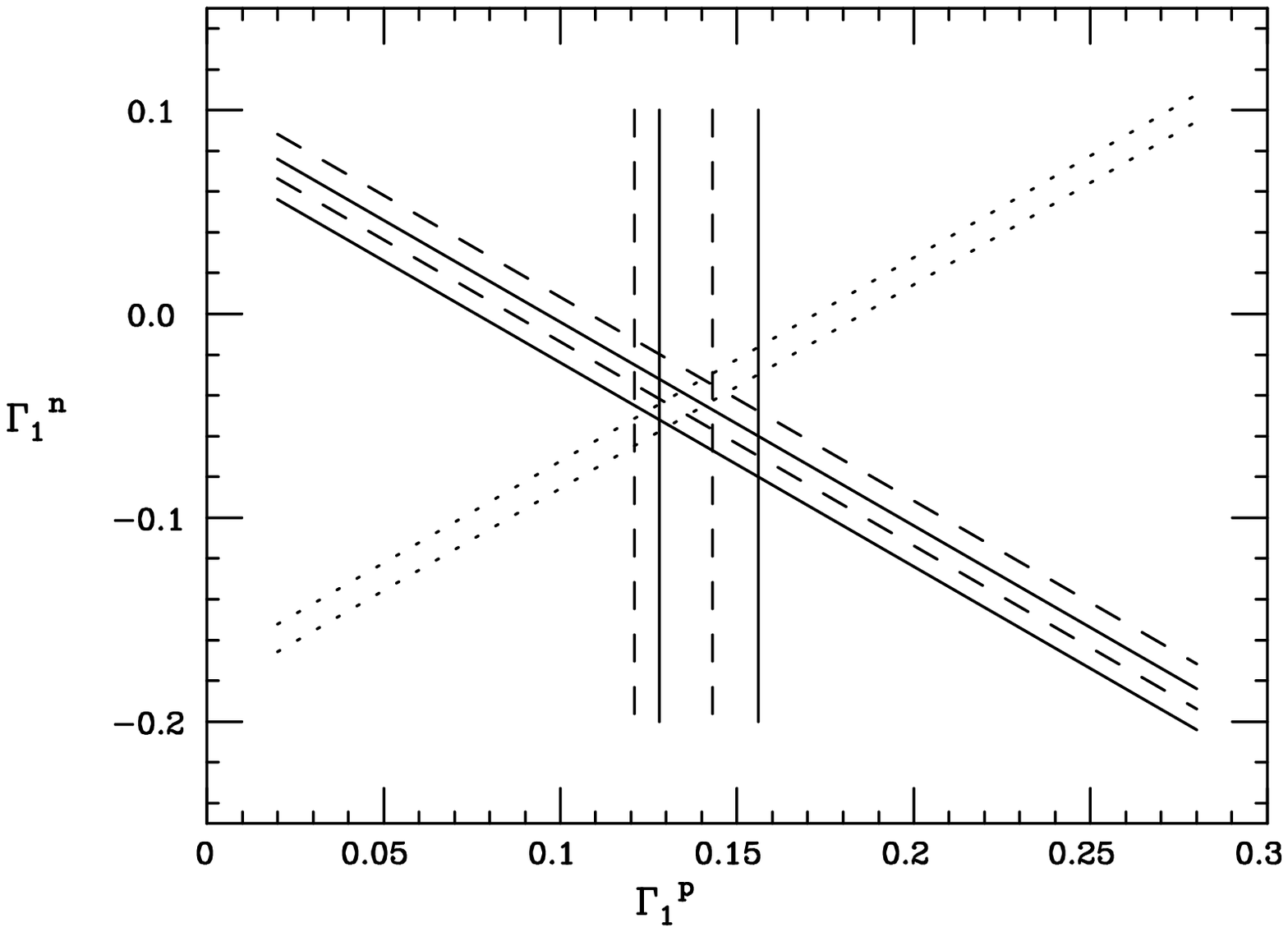}\hfil\vskip-4.2truecm
\bigskip\noindent{\abstractfont\baselineskip6pt\narrower
Figure 2: Values of the first moment $\Gamma_1$ eq.~\fmom\ from
table~1 evolved to the common scale of $Q^2=5$~GeV$^2$.
The solid lines correspond to the SMC experiments and the dashed
lines to E143 experiments.
The dotted line
is the prediction of the Bjorken sum rule eq.~\bjsr (as the solid
lines in fig.~1).
\smallskip}
\medskip
\endinsert
\medskip
\subsec{The ``spin crisis'' and the anomaly}
\smallskip
The value of the singlet axial charge thus determined
has attracted a good deal
of attention because of its smallness. Small here
refers to the expectation, based on the Zweig rule,
that $a_0\approx a_8$,
which the value in eq.~\guisig\ violates by several standard deviations.
However, $a_8$ is scale-independent, while $a_0$ depends on $Q^2$ according to
eq.~\azevol: this raises immediately the question of the scale
at which these two quantities should be compared. Indeed, the Zweig
rule is a phenomenological rule of the naive quark model. Matrix elements of
operators measured in hard processes should agree with the predictions
of this model at some typical low hadronic scale: it is however clear that,
at a sufficiently low scale, $a_0(Q^2)$ [eq.~\azevol]
grows as large as desired; in particular, it is of the same
size as $a_8$ around $0.5$~GeV$^2$.
Whereas clearly perturbative evolution
at such low scales cannot be trusted, nonperturbative
estimates\ref\bal{R.~D.~Ball. {\it Phys}. {\it Lett}. {\bf B266}
(1991) 473.} suggest that this scale dependence may actually be even
stronger than perturbatively expected.

More specifically, one may take
the point of view that naive constituent quark observables are related
to parton distributions measured in hard processes by a (generally
$x$-- and $Q^2$--dependent) renormalization. A simple way of modelling this is
to assume that the nucleon structure functions are given by combinations
of constituent quark structure functions, each of which in turn is
the convolution of a parton density inside the constituent quarks
times the constituent quark density in the nucleon\nref\convmod{G.~Altarelli
et al., {\it Nucl}. {\it Phys}. {\bf B69} (1974) 531.}\refs{\alrid, \convmod}.
The
constituent quark densities are then
scale independent but target-dependent, while
the parton densities satisfy the
Altarelli-Parisi equations and should be universal. The model is tested by
verifying this universality (for instance, it allows predicting
the second moment
of the pion parton distributions in terms of the proton ones)
and seems to be in good agreement with
experimental data\ref\gui\convtest{G.~Altarelli, S.~Petrarca and F.~Rapuano,
preprint CERN-TH/95-273, {\tt hep-ph/9510346}.}. The violation of the
Zweig rule, which is then  by construction  satisfied by
constituent quarks, is
attributed to their partonic substructure: the apparent contradiction
with the quark model is thus removed, but the observed
effect is not explained.

A deeper understanding of the meaning of
 observed violation of the
Zweig rule, and more in general of the first moment of $g_1$, can be obtained
in the QCD parton model.
Matrix elements of operators are endowed with a partonic interpretation by
identifying them with moments of parton distributions. The definition
of the operators is however ambiguous since a renormalized operator can
always be redefined by a finite renormalization:
given a specific definition of, say, the isotriplet axial current
${j^\mu_5}_{I=1}$, we can
define a new current
\eqn\renor{{j^\mu_5}_{I=1}^\prime=Z(\alpha_s){j^\mu_5}_{I=1},}
where
$Z[\alpha_s(Q^2)]=1+Z^{(1)} \alpha_s(Q^2)+\dots$. The relation of the current
to physical observables
will be modified accordingly: for example if $\Gamma_1^{I=1}=C_{\rm NS}
(Q^2) \langle
{j^\mu_5}_{I=1}\rangle$, then $\Gamma_1^{I=1}=C_{\rm NS}(Q^2)Z^{-1}(Q^2)
\langle
{j^\mu_5}_{I=1}^\prime\rangle$. This is, a priori, as good a definition of the
axial current as the original one, and the ensuing ambiguity
(factorization scheme ambiguity) only goes away at $Q^2\to\infty$,
or to all perturbative orders. It may, however, be fixed
by physical requirements, such as the preservation of physical symmetries.
In particular, if a current is conserved (as
the  nonsinglet axial current) it is protected against scale dependence:
because $\partial_\mu {j^\mu_5}^{I=1}=0$ then necessarily
$Q^2\smallfrac {d}{d Q^2} \partial_\mu {j^\mu_5}^{I=1}=0$, hence $j^\mu_5$
(as a renormalized composite operator) does
not depend on $Q^2$. It follows that
the redefinition of eq.~\renor\ is not allowed
if one imposes that chiral symmetry be preserved, because it would spoil
this scale independence, and there exists a ``natural''
normalization of the current, namely, that which preserves
chiral symmetry.

The current  matrix element can then be identified
with the first moment of the polarized quark distribution:
for instance the isotriplet charge is
\eqn\nsingp{
a_{I=1}=g_A=\int_0^1\!dx\, \Delta q_{I=1}(x)=
\Delta u(1, Q^2)-\Delta d(1, Q^2),}
where $\Delta u(x,Q^2)$, $\Delta d(x,Q^2)$ are
the polarized up and down distributions,
whose $N$-th moment is given
by
\eqn\mom{\Delta q_i(N,Q^2)=\int_0^1\!dx\,x^{N-1}\Delta q_i(x,Q^2);}
and similarly for other nonsinglet charges.
The conserved charge associated to ${j^\mu_5}_{I=1}$
is the total (isotriplet) quark
helicity, so $a_{I=1}$ can be interpreted as the total (isotriplet) quark
polarization, in agreement with naive partonic expectation.

Things are more complicated in the singlet case; first,
because there now  are two parton distributions rather
than one, i.e. the gluon distribution $\Delta g(x,Q^2)$
and the singlet quark distribution
\eqn\sing
{\Delta \Sigma(x,Q^2)=\sum_{i=1}^{n_f} \Delta q_i(x,Q^2),}
where $\Delta q_i(x,Q^2)$ is the quark distribution of flavor $i$
and $n_f$ is the number of flavors activated at the scale $Q^2$;
and furthermore, because the
current is no longer conserved, thus as a renormalized operator it
depends on scale, and there
is  no unique ``natural'' way to normalize
the quark singlet and gluon first moments, which will mix upon evolution.

Indeed, the classical conservation of the singlet axial current
is spoiled at the quantum level by the axial anomaly\ref\roman{See
R.~Jackiw, in S.~Treiman, R.~Jackiw,
B.~Zumino
and E.~Witten, {\it ``Current Algebra and Anomalies''} (World Scientific,
Singapore, 1985) for a review.}:
\eqn\anom
{\partial_\mu j^\mu_5=n_f{\alpha_s\over2\pi}\tr\epsilon^{\mu\nu\rho\sigma}
F_{\mu\nu}F_{\rho\sigma}.}
The singlet current is thus unprotected and
requires an infinite renormalization, which
induces a multiplicative
scale dependence of the renormalized current of the form\ref\adbar{S.~Adler
and W.~Bardeen, \PR\vyp{182}{1969}{1517}.}
\eqn\scalda{{d\over d t}  j^\mu_5=\gamma_5(\alpha_s) j^\mu_5,}
where $t\equiv\ln(Q^2/\Lambda^2)$.
The anomalous dimension $\gamma_5(\alpha_s)=\gamma_5^{(2)}\alpha_s^2+\dots$
starts at two loops (because
the nonconservation of the current is a one--loop effect) and
depends on the specific choice
of operator normalization; the scale dependence of $a_0$ induced by
it was given in the  \MS\ scheme in eq.~\azevol.

Once this choice is made,
however, the remaining normalization ambiguities are fixed by the
anomaly equation.
Indeed, eq.~\anom\ holds as an equation between renormalized composite
operators\ref\mich{M.~Bos, \NP \vyp{B404}{1993}{215}.}, and it therefore
implies
that the operator
$\tr\epsilon^{\mu\nu\rho\sigma}
F_{\mu\nu}F_{\rho\sigma}$ must mix with the current in such a way
as to compensate (to all orders in perturbation theory) its scale dependence:
\eqn\RGanom{{{d}\over
{d t}}n_f{\alpha_s\over2\pi}\tr\epsilon^{\mu\nu\rho\sigma}
F_{\mu\nu}F_{\rho\sigma}=\gamma_5(\alpha_s)\partial_\mu j^\mu_5.}
This condition fixes the normalization of the singlet gluon operator that
mixes with the axial current, as well as
the coefficient of this mixing --- that
there exists a scheme where eq.~\RGanom\
is satisfied is the content of the Adler--Bardeen
theorem\cite\adbar\ and modern versions thereof\cite\mich.

However, there is  still an obstacle in the identification of matrix elements
of operators with partonic quantities: namely,
we cannot simply
identify the first moments $\Delta \Sigma(1,Q^2)$
and $\Delta g(1,Q^2)$ of the quark and gluon distribution
[defined in analogy to eq.~\mom]
with matrix elements of two leading twist operators, because
there is only one local
leading
twist (i.e. twist 2) operator with spin corresponding to
the first moment of $g_1$ (i.e. spin 1).
It follows that while it is natural to identify
$\Delta g(1,Q^2)$  (up to an overall constant, still
arbitrary at this stage),
with the quantity that mixes with the matrix element $a_0$
of $j^\mu_5$ according to
eq.~\RGanom, i.e.
\eqn\gluRG{
-{{d}\over
{d t}}n_f{\alpha_s\over2\pi}\Delta g(1, Q^2)
=\gamma_5(\alpha_s) a_0(Q^2),}
there is no unique way to identify $a_0(Q^2)$ itself. Indeed,
if the quark singlet and gluon
anomalous dimensions are computed for higher moments, where a tower
of quark and gluon operators (for all odd $n\ge3$) are naturally
identified with the corresponding moments of quark and gluon
distributions\ref\koda{J.~Kodaira et al., \PR\vyp{D20}{1979}{627};
\NP\vyp{B159}{1979}{99}; J.~Kodaira, \NP\vyp{B165}{1979}{129}.},
and then analytically continued to the first moment,
the result depends on the adopted regularization:
if dimensional regularization is used throughout\cite\merner, then the
anomalous dimension of $\Delta \Sigma(1,Q^2)$ coincides
with $\gamma_5(\alpha_s)$; if instead infrared collinear singularities
are regulated by means of an explicit
regulator\cite\koda\
(such as putting
the incoming particle off-shell) then the anomalous dimension
of $\Delta \Sigma(1,Q^2)=\Delta \Sigma(1)$ vanishes. Note
that in either case
eq.~\gluRG\ is satisfied, thereby fixing the constant of proportionality
between $\Delta g(1, Q^2)$ and the operator which mixes with $a_0$ . These two
results
correspond, respectively, to identifying the first moment of
the quark distribution with the two eigenvectors of perturbative
evolution: in the former case the singlet quark distribution
is identified as
$\Delta \Sigma(1,Q^2)=
a_0(Q^2)$, and in the latter case it is identified with the scale-invariant
combination $\Delta\Sigma(1)=a_0(Q^2)+n_f{\alpha_s\over2\pi}\Delta g(1, Q^2)$.

This ambiguity can be pinned down only on the basis of physical requirements
on parton distributions. Now, it may be shown\ref\fact{R.~D.~Carlitz,
J.~C.~Collins and A.~H.~Mueller, \PL \vyp{B214}{1988}{229};
G.~Altarelli and B.~Lampe, \ZP \vyp{C47}{1990}{315};
W.~Vogelsang, \ZP \vyp{C50}{1991}{275}.}
 that, in schemes where
$a_0=\Sigma(1,Q^2)$, soft contributions are
partly included in the hard coefficient function, rather than
being properly factorized in the parton distributions:
this explains why these
contributions are removed by an infrared regulator, which instead
yields schemes where the quark
is identified with the conserved eigenvector.
Furthermore, in the latter schemes  $\Delta \Sigma(1)$
can be directly shown\ref\sfpol{S.~Forte,
\NP\vyp{B331}{1990}{1}.} to coincide with the nucleon matrix element
of the canonical, conserved quark helicity operator,
the scale-dependence of $a_0$ then being  due to a contribution
to it from  particle creation induced by the
axial anomaly. In these schemes the Zweig rule expectation acquires a
precise meaning\ref\alros{G.~Altarelli and G.~G.~Ross,
\PL\vyp{B212}{1988}{391}.}, since it predicts the approximate equality of
two scale--independent quantities: $a_8\approx\Delta \Sigma(1)$.
The (scale--dependent) deviation of $a_0$
from this value would then be  explained by the gluon contribution to
it\cite\alros.
Of  course, whether this is actually the case rests
with experiment: we will discuss this in sect.~4.2.

The choice of factorization scheme in the definition of polarized
quark and gluon distribution, besides affecting the physical
interpretation of the singlet first moment of $g_1$, has a
significant effect in a finite-order computation of the
scale dependence of $\Gamma_1$, and may thus substantially
affect the extraction of this quantity from the data. Indeed,
in the class of schemes where $\Delta \Sigma(1)$ is scale-independent,
the singlet $\Gamma_1$ is given by\eqnn\singao\eqnn\singfo
$$\eqalignno{\Gamma_1^{\rm sing.}&={1\over2} \langle e^2\rangle C_{\rm S}
(Q^2)\left(\Delta \Sigma (1)-{\alpha_s\over 2\pi}n_f
\Delta g(1,Q^2)\right)&\singao \cr
&={\langle e^2\rangle\over 2}\left[ C_q(Q^2)
\Delta \Sigma(1)+ 2 n_fC_g(Q^2)\Delta
g(1,Q^2)\right],&\singfo\cr}$$
where the quark and gluon coefficient functions are given by
\eqn\qgcf{C_g(Q^2)=-{\alpha_s\over 4\pi} C_q(Q^2)=
-{\alpha_s\over 4\pi}C_{\rm S}(Q^2)}
to all orders in perturbation theory. This is to be
contrasted to schemes where $\Delta \Sigma(1,Q^2)=a_0(Q^2)$, so
$C_g(Q^2)=0$.

If
the coefficient functions are determined at any finite
perturbative order, eq.~\singao\ implies  that $a_0(Q^2)$ is not simply found,
as in sect.~3.1,
dividing $\Gamma_1^{\rm sing.}$ by the coefficient function $C_{\rm S}(Q^2)$:
indeed,
using the  expression of $C_q$ and $C_g$ to order $\alpha_s^k$
in eq.~\singfo\ one gets
\eqn\fmomerr{\smallfrac{2}{C_{\rm S}(Q^2)\langle e^2\rangle}
\Gamma_1^{\rm sing.}-a_0(Q^2)=
-  n_f\left(\smallfrac{C_{\rm S}^{(k)}}
{2\pi}\right)\alpha_s^{k+1} \Delta g(1,Q
^2)+
O(\alpha_s^{k+2}),}
where $C_{\rm S}^{(k)}$ is the coefficient of $\alpha_s^k$ in the
perturbative expansion of $C_{\rm S}(Q^2)$.
Thus all determinations of $a_0$ discussed so far are affected by an
error of this size, and could only be improved by extracting
$\Delta \Sigma(1)$ and $\Delta g(1,Q^2)$ directly from the data
(for example fitting eq.~\singfo\ to the data) and then computing
$a_0(Q^2)=\left(\Delta \Sigma(1)-{\alpha\over 2\pi}
2n_f\Delta g(1,Q^2)\right) $.
A more conservative point of view is
that this is an intrinsic and unavoidable ambiguity of the computation.
The size of the ambiguity of course depends crucially on the size
of the gluon and decreases rapidly as the scale and the perturbative order
increase. At order $\alpha_s$, assuming that the quark respects
the Zweig rule and the gluon makes up for the difference, and taking
the value of $a_0$ from table~1, the correction at 3~GeV$^2$ is around
15\% of the value if $a_0\approx 0.5$, and around 50\% if $a_0\approx 0.15$.
Note that this uncertainty is not included
in the errors given in table~1.

The scale dependence of $a_0$ induced by the axial anomaly is
determined perturbatively by the anomalous dimension $\gamma_5(\alpha_s)$,
which is a universal (i.e. target--independent) property of the axial
current. It has been shown recently\ref\shove{G.M.~Shore and G.~Veneziano,
\PL\vyp{B244}{1990}{75}; \NP\vyp{B381}{1992}{23}; S.~Narison,
G.M.~Shore and G.~Veneziano, \NP\vyp{B433}{1995}{209}.} that this universality
actually persists beyond perturbation theory: the
singlet axial charge satisfies a Goldberger-Treiman equation which relates
it to the decay constant $F_{\Phi}$
and irreducible coupling to the nucleon $g_{\Phi\bar BB}$
of a singlet pseudoscalar state $\Phi$
(defined in analogy to the $\pi$ coupling and decay constant
in the usual Goldberger--Treiman relation): $a_0(Q^2)=
g_{\Phi\bar BB}\,F_\Phi(Q^2)$.  This pseudoscalar meson
is however not a physical state of the theory due to the presence
of the anomaly in this channel, so these couplings could only be
measured on the lattice. However, the
decay constant can be expressed in terms of the topological
susceptibility, which is a property of the QCD vacuum. Furthermore,
all the scale dependence of $a_0$ comes from this quantity, and  is thus
contained in a universal, target-independent factor. It follows
that if the explanation of the observed Zweig rule violation
is in this scale dependence, i.e. in the smallness
of the scale-dependent term $F_\Phi$, the effect will be universal,
and not specific of the nucleon. It is interesting to compare
this possibility with an alternative proposal\ref\sfinst{S.~Forte and
E.~V.~Shuryak, {\it Nucl. Phys.} {\bf B357} (1991) 153.}
where the effect is instead related to a small value of the coupling
$g_{\Phi\bar BB}$ due to instantons. The effect would then be
strongly target dependent, and have a different scale dependence.
More  data could thus shed light on our understanding of
the QCD vacuum beyond perturbation theory.
\bigskip
\newsec{The structure function in the $(x,Q^2)$ plane}
\medskip
We have seen in the previous sections that the scale dependence
of $\Gamma_1$ [eq.~\gamdec] driven by the Wilson coefficient functions
as well as by the anomalous dimension of the singlet axial
current is quite large, that it substantially affects the extraction of
current matrix elements from the data, and that it is actually responsible for
a large part of the uncertainty on their determination.
This suggests that a detailed understanding
of the evolution of the full structure function
$g_1$ in the $(x,Q^2)$ plane is necessary in order to accurately
assess
this uncertainty, and pin it down as much as possible. Indeed,
experimental data are taken within a limited range in $x$
($0.003\le x\le0.7$ for SMC and $0.03\le x\le 0.75$ for E143)
and at different scales in each $x$ bin, with the lowest
$x$ points being taken at low $Q^2$ and conversely
($1.3\le Q^2\le48.7$~GeV$^2$ for SMC and $1.3\le Q^2\le 9.2$~GeV$^2$
for E143). The determination of moments of $g_1$ requires thus both
extrapolation in $x$ and evolution to a common scale. The two
problems are closely related because $g_1(x,Q^2)$ is
determined in terms of polarized parton distributions
\eqn\gone
{g_1(x,Q^2)=
\smallfrac{1}{2}\left[
C_{\rm NS}\otimes\Delta q_{\rm NS}
+ \langle e^2\rangle\left(C_{\rm S}\otimes \Delta\Sigma
+ 2n_f C_g\otimes\Delta g\right)\right],}
(where
$\otimes$ denotes the usual convolution with respect to
$x$). These evolve perturbatively
according
to the Altarelli--Parisi equations,
so  parton distributions at $(x_0,Q_0)$ are causally
determined from their values at $x>x_0$, $Q<Q_0$.

The values of the first moments
in table~1 are arrived at by assuming  the
scattering asymmetry $A_1$ to be scale--independent, i.e. by approximately
evolving the data to a common scale on the assumption that the scale
dependence of the polarized structure function $g_1$ is the
same as that of the unpolarized one $F_2$, and  extrapolating
to small (and large) $x$ by fitting a phenomenological shape to the
last few data points.
In order to critically examine these assumptions, we must study the small $x$
behavior and scale dependence of $g_1$ in perturbative QCD.
\medskip
\subsec{The small $x$ behavior of polarized parton distributions}
\smallskip
Whereas the extrapolation of the structure function $g_1$ from
the measured range to $x=1$ is under theoretical control and
amounts to a small uncertainty, the extrapolation to small $x$
is more subtle. This is due to the fact that $g_1$ must vanish
identically for kinematic reasons at $x=1$; moreover, the form
of its drop is already constrained by available data and agrees
with expectations based on QCD counting rules~\ref\brod{See
S.~J.~Brodsky, M.~Burkardt and I.~Schmidt, {\it Nucl. Phys.} {\bf B441} (1995)
197 and references therein.}. On the contrary, the observed small $x$
behavior has thwarted several times theoretical prejudice. In fact,
at least in principle, the associated
uncertainty is infinite, since
the $x\to0$ limit can never be attained (it corresponds
to infinite energy) and, because
perturbative evolution
proceeds from larger to smaller $x$, it is a
priori impossible to exclude a growth in the unexplored small $x$ region.

The traditional expectation for the small $x$ behavior of structure
functions is
embodied in Regge
theory,
which at some low scale should provide the small $x$
behavior of parton distributions
that are input to perturbative evolution .
Dominance of Regge poles
would predict\ref\heim{R.~L.~Heimann, {\it Nucl. Phys.} {\bf B64} (1973)
429.} that both the singlet and nonsinglet contributions
to $g_1$ decrease or are at most constant as $x\to0$. Notice that
in the unpolarized case  Regge theory
would have the singlet contribution to $F_2$  behave like a constant
(or perhaps grow slightly, for a supercritical Pomeron)
 and the nonsinglet drop as $\sqrt{x}$; both
the nonsinglet\ref\NMC{NMC Collaboration, M.~Arneodo et al., \PR\vyp{D50}
{1994}{1}.} and
singlet\ref\DAS{R.~D.~Ball and S.~Forte, \PL\vyp{B335}{1994}{77}.}
predictions are in excellent agreement with available experimental data.
Because in LO
$F_2\propto x q(x)$, while $g_1\propto \Delta q(x)$, the Regge expectations
for the singlet can be made consistent with the expectation\cite\brod\
(partly based on QCD) that ${\Delta q\over q}\neath{\sim}{x\to0}x$,
if one assumes  that both $F_2$ and $g_1$ are constant or almost
constant at small $x$ at a low scale.
  The SMC data
on $g_1^p$, however, seem to indicate a rapid growth of $g_1$ as $x\to0$
(see fig.~3a). One might have thought this to be due to a stronger
increase of the singlet at small $x$, perhaps due to Regge cuts, but this
is  belied by the SMC data on $g_1^d$,
which should then look like the proton in the small $x$ region, whereas
in actual fact it is negative there (fig.~3b).

Now, even if at some low scale parton distributions behave according
to Regge expectations, they will be modified by perturbative evolution.
In particular, it has been known for long that LO
perturbative evolution
eventually leads to a rise of $g_1$ at small
$x$\nref\polscal{M.~A.~Ahmed and
G.~G.~Ross, {\it Phys. Lett.} {\bf B56} (1975)
385; M.~B.~Einhorn and J.~Soffer, {\it Nucl. Phys.} {\bf B74} (1986) 714;
A.~Berera, {\it Phys. Lett.} {\bf B293} (1992)
445.}\nref\BFR{R.~D.~Ball,
S.~Forte and G.~Ridolfi, {\it Nucl}. {\it Phys}.
{\bf B 444} (1995) 287}\refs{\polscal,\BFR}
of the same form as the rise of the unpolarized structure
function $F_2$\ref\DLLA{A.~De~R\'ujula et al., \PR\vyp{D10}{1974}{1649}.};
the main difference is  that whereas
$F_2$ is positive--definite $g_1$ is not,
so the rise may correspond to $g_1$ growing either
large and positive or large and negative.
A closer look reveals both analogies and differences between
the polarized and unpolarized cases\cite\BFR.
The small $x$ behavior of parton distributions is dominated by the
rightmost singularity of perturbative
anomalous dimensions in the space of moments.
In the polarized case, this singularity is located at $N=0$ for
the nonsinglet as well as for
all entries in the matrix of singlet anomalous dimensions,
and it is already present  at LO.
In the unpolarized case it is located at $N=1$ for
the nonsinglet and at $N=0$ for the singlet; in the latter case
the singularity is present at LO in
the gluon anomalous dimensions, but only starts at NLO
in the quark ones
[defining the moment variable $N$ according to eq.~\mom].

As a consequence, both the singlet and nonsinglet polarized distributions
grow according to\cite\BFR
\eqn\smxb{\Delta f(x,Q^2)\sim
{1\over\sqrt{\sigma}}e^{2\gamma_f\sigma},}
where $\sigma\equiv\sqrt{\xi\zeta}$, $\rho\equiv\sqrt{\xi/\zeta}$,
$\xi\equiv\ln\smallfrac{x_0}{x}$,
$\zeta\equiv\ln\smallfrac{\alpha_s(Q_0^2)}{\alpha_s(Q^2)}$,
and $x_0$ is a reference value of
$x$ such that the approximate small $x$ form of the anomalous dimensions
is applicable for $x\lsim x_0$ and $Q^2\gsim Q^2_0$.
In the nonsinglet case, $\Delta f=\Delta q_{\rm NS}$ and
\eqn\gamns
{\gamma_{\rm NS}^2= {8\over 33- 2 n_f}.}
In the singlet case the quantities
that  display this growth are the linear combinations of quark
and gluon $v^\pm=\Delta \Sigma+C^\pm \Delta g$, with
\eqn\evecc
{C^\pm=2\left(1\pm\sqrt{1-{3n_f\over32}}\right);
\qquad
\gamma_{\pm}^2= \gamma_{\rm NS}^2\left(5\pm4\sqrt{1-{3n_f\over 32}}\right).}
Notice
that this is a scheme independent result.
The behavior of the unpolarized nonsinglet is the same
as that of the polarized singlet (but with different $\gamma$),
while the behavior of the unpolarized singlet differs, first
because the quantity which grows according to eq.~\smxb\
is $x g(x,Q^2) $ rather than $\Delta g(x,Q^2)$,
 and furthermore because the growth is driven by the gluon, i.e.
the unpolarized quark is proportional
to the unpolarized gluon (and of order $\alpha_s$ with respect to it).

This implies that eventually the singlet distributions indeed dominate
$g_1$ at small $x$. In both singlet eigenvectors the quark
and gluon have opposite sign, and asymptotically $\Delta \Sigma >0$
if $\Delta \Sigma(x_0,Q_0)>- C^+\Delta g(x_0,Q_0)$: hence,
for most plausible starting quark and gluon $\Delta \Sigma$
grows large and negative and $\Delta g$ large and positive. As a consequence,
asymptotically $g_1$ at small $x$ must be large and negative.
The nonsinglet distribution however also grows asymptotically;
the growth has the same form as for the singlet distributions, only with
a smaller value of $\gamma_f$, its sign is the same as that of the
starting distribution. Therefore, the value of $x$ where the singlet
actually does dominate could  in practice be extremely small. This is
to be contrasted to the unpolarized case, where the growth of eq.~\smxb\
is down by a power of $x$ in the nonsinglet compared to the singlet,
so that the singlet already dominates around $x\sim10^{-3}$.

Of course, the perturbatively generated growth of eq.~\smxb\ is
only visible provided it dominates over a possible growth of the
starting parton distributions:
even though Regge theory does not favor such a growth,
it cannot be firmly ruled out. For instance, if parton distributions
rise as $\Delta f(x,Q_0)\sim
x^{-\lambda}$, then the evolved distributions will
behave according to eq.~\smxb\ only for
$\rho\lsim{\gamma_f\over\lambda}$, but for smaller values of $x$ they
will behave as\cite\BFR\
\eqn\hardasymp{\Delta f\sim x^{-\lambda}
\left({\alpha_s(Q_0^2)\over\alpha_s(Q^2)}\right)^{\gamma^2_f\over\lambda},}
i.e. they  reproduce
the boundary condition up to an $x$--independent correction;
in the singlet case if the starting
quark and gluon have different small $x$--behavior the most singular
one dominates asymptotically.
 In such case, the onset of the asymptotic
small $x$
behavior of eq.~\hardasymp\ will be yet
slower\ref\stirsx{T.~Gehrmann and W.~J.~Stirling, Durham preprint
DTP/95/62, {\tt hep-ph/9507332}.}: firstly, because the
small $x$ solution to the evolution equations is then dominated by
the singularity of the boundary condition rather than the
singularity at $N=0$ in the anomalous dimension, therefore,  approximating the
anomalous dimension with this singularity is less accurate; also,
the two singlet eigenvectors only mix due to the $x$-independent
correction in eq.~\hardasymp, thus the dominance of the most singular
behavior only sets in rather slowly.

It is interesting to see how these LO predictions
are modified at higher orders. At NLO
all polarized anomalous dimensions have a $1\over N^3$ singularity.
This leads to a correction to the LO
of the form\cite\BFRb
\eqn\smxbNLO{\Delta f^{\rm NLO}(x,Q^2)=
\left[1+\epsilon_f
\big(\smallfrac{\rho}{\gamma_f}\big)^3
\left({\alpha_s(Q^2_0)}-{\alpha_s(Q^2)}\right)\right]\Delta f^{\rm LO}(x,Q^2),}
where $\Delta f^{\rm LO}(x,Q^2)$ is given by eq.~\smxb, and
the coefficients
$\epsilon$ are explicitly
\eqn\delco{\epsilon_{\rm NS}= \smallfrac{8}{3\pi\beta_0};\qquad
\epsilon_\pm =\smallfrac{112}{3\pi\beta_0}\bigg[
(1-\smallfrac{n_f}{14})\pm
\smallfrac{13}{14}(1-\smallfrac{11n_f}{104})
\Big/\sqrt{1-\smallfrac{3n_f}{32}}\bigg],}
corresponding to a further rise proportional to $\xi^{3/2}$ of the
parton distributions at small $x$, with coefficient
${\epsilon_{\rm NS}/\gamma_{\rm NS}^3}\approx
{\epsilon_+/\gamma_+^3}\approx\half$.
 Notice that the form of the
small $x$ eigenvectors is unaltered, and that these results
are scheme--independent.

\topinsert
\vskip-3.5truecm\vbox{\hbox{\hskip-1truecm
\hfil\epsfxsize=10truecm\epsfbox{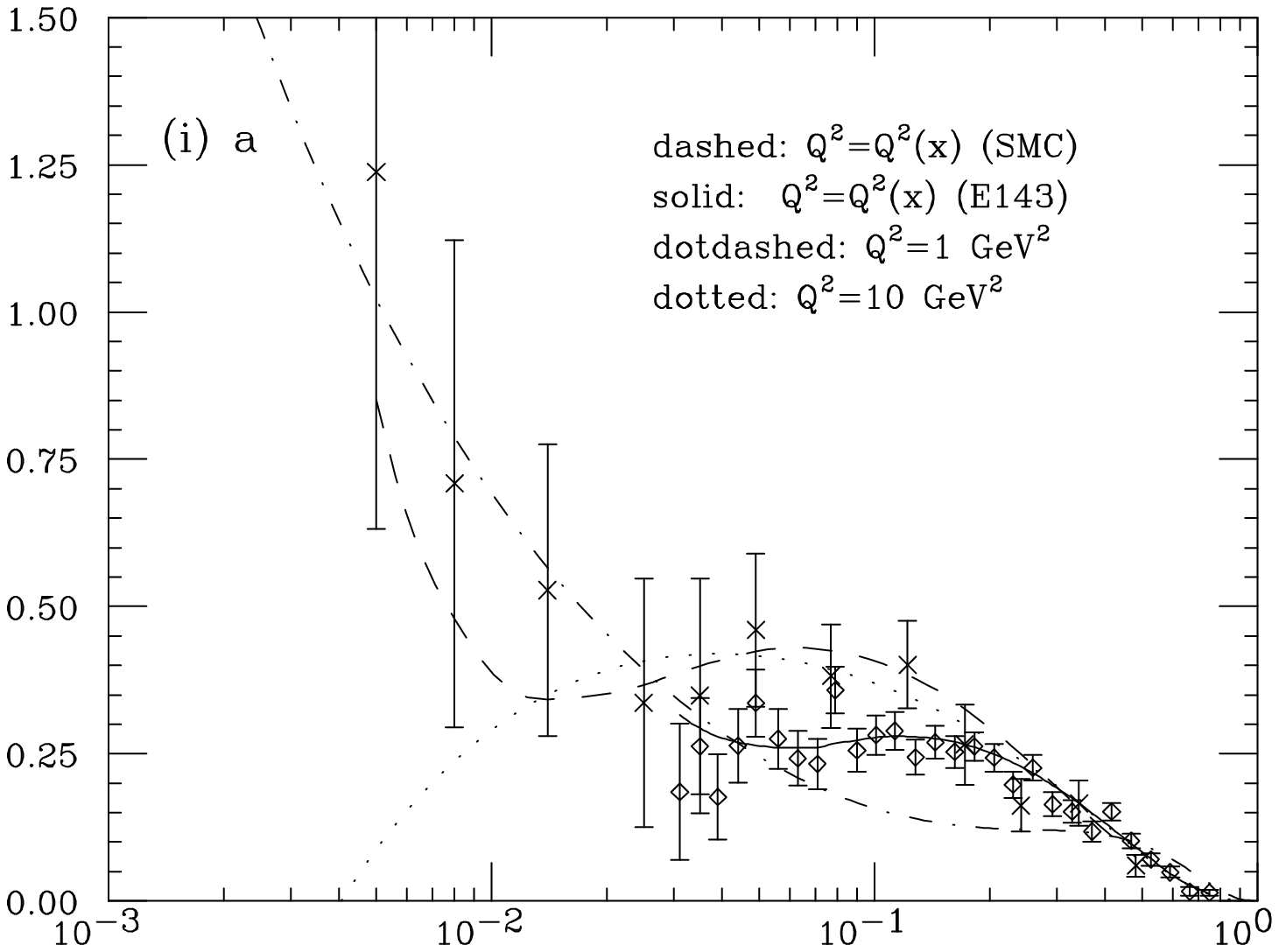}\hskip-2.truecm
\epsfxsize=10truecm\epsfbox{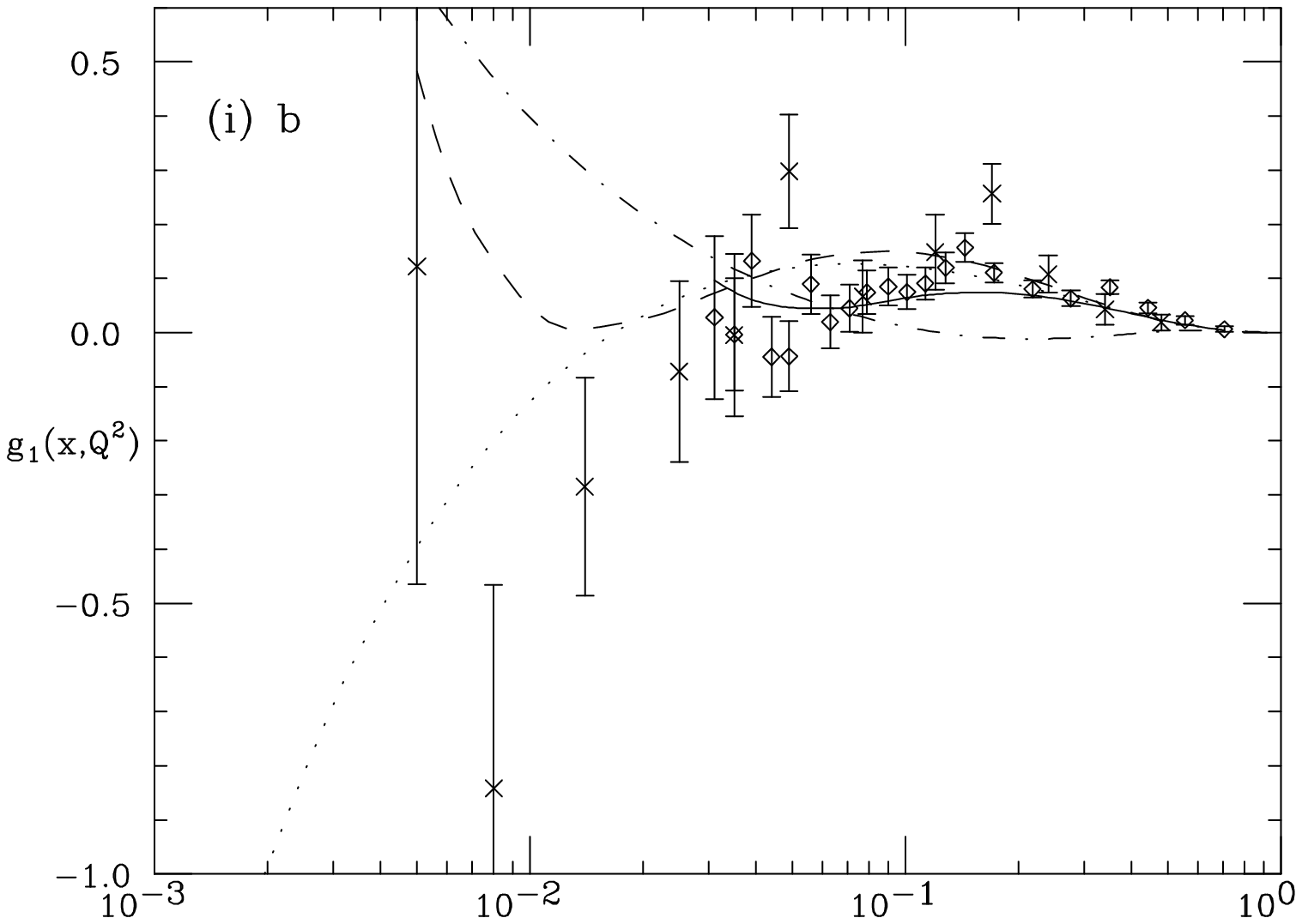}\hfil}
\vskip-7.5truecm
\hbox{\hskip-1truecm
\hfil\epsfxsize=10truecm\epsfbox{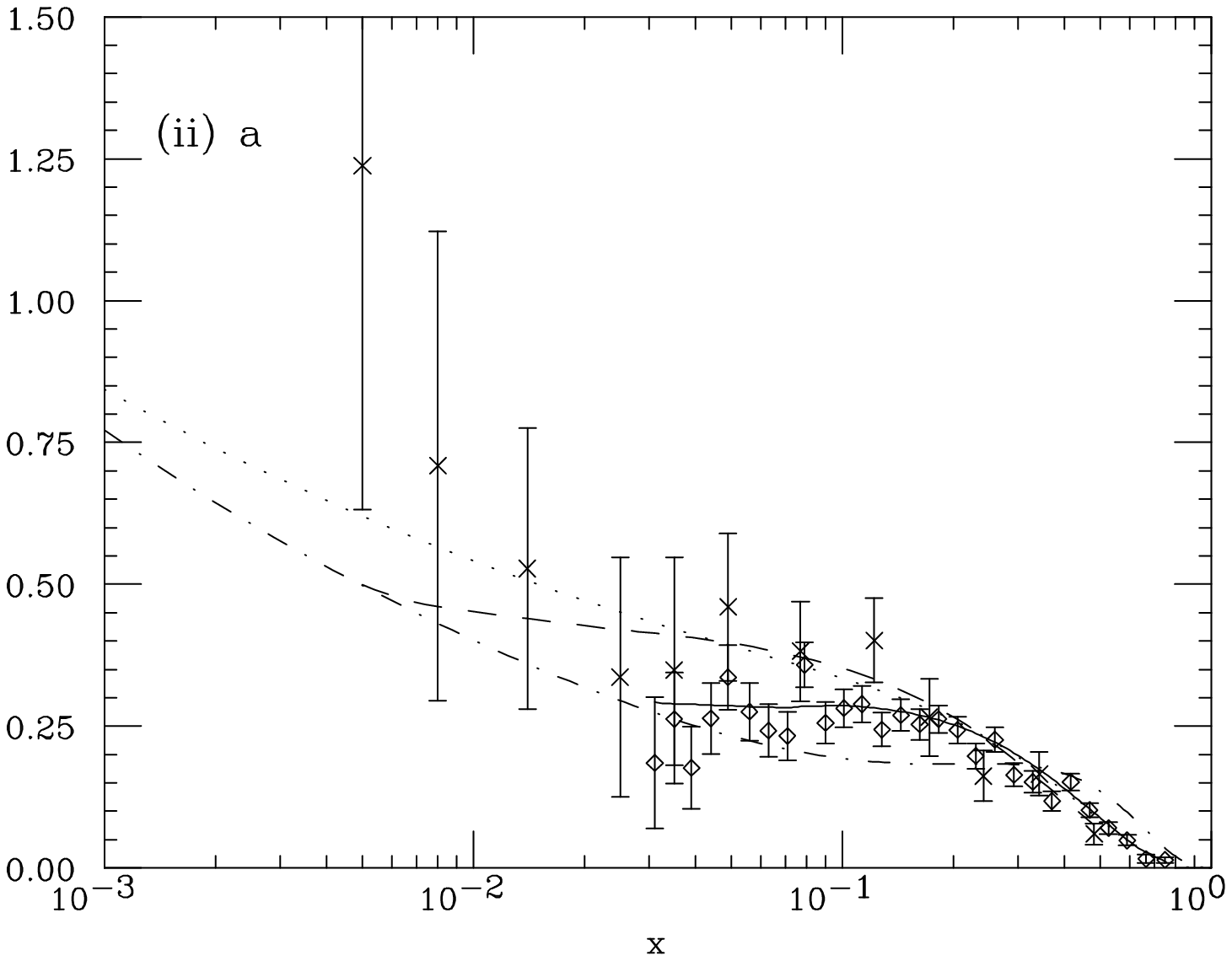}\hskip-2.truecm
\epsfxsize=10truecm\epsfbox{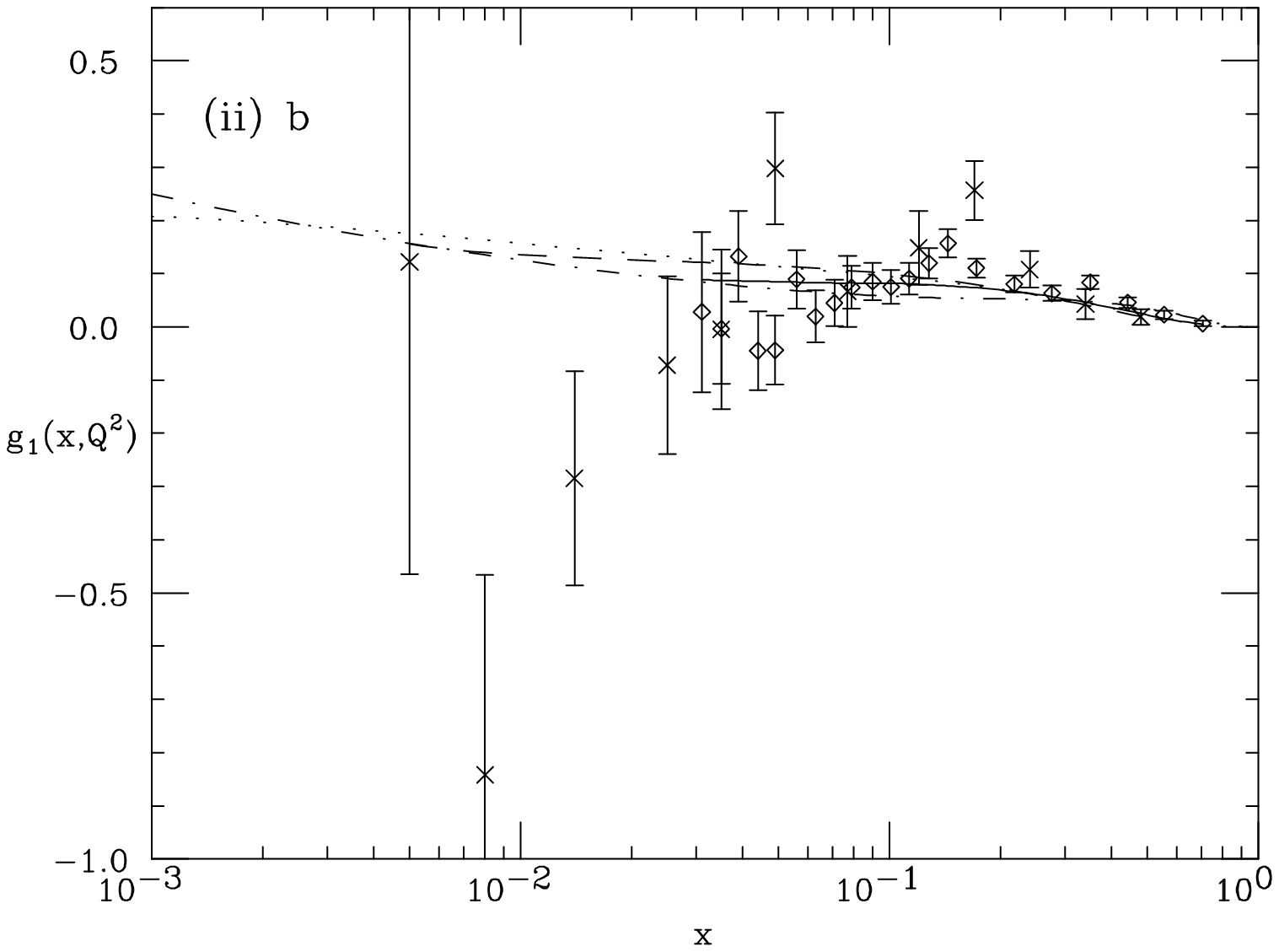}\hfil}
\vskip-4.truecm
\bigskip\noindent{\abstractfont\baselineskip6pt\narrower
Figure 3: Plot of  $g_1(x)$ for (a) proton and (b) deuteron.
The crosses are SMC data and the diamonds E143 data.
The curves correspond to a fit\cite\BFR to the proton data only
with  valence-like nonsinglet and
(i) ``maximal'' gluon and steep input quark singlet
or (ii) ``minimal'' gluon and flat input quark singlet. }}
\medskip
\endinsert
At NLO the Mellin transforms of the quark singlet and nonsinglet
and gluon coefficient functions also acquire a singularity
at $N=0$,
which is however only of order $1\over N^2$ and therefore does
not contribute to the asymptotic small $x$ behavior eq.~\smxb.
This singularity corresponds to a $\ln {1\over x}$ rise
(with positive sign)
of the functions $C(x,Q^2)$ in eq.~\gone. At small $x$, this  leads to
a rise of $g_1$ even when the various parton distributions are flat.
Even
though this rise is weaker than that induced by perturbative
evolution, it could dominate at low scale
where evolution effects are small; its precise form  is
scheme--dependent, however.

At yet higher orders, the expected generic behavior
of anomalous dimensions is $\gamma\neath\sim{N\to0}(\alpha_s^n/N^{2n-1})$.
The
small $N$ behavior of the NLO
coefficient functions suggests that these might instead behave
as  $C\neath\sim{N\to0}\left(
\alpha_s/ N^2\right)^{n-1}$, which would guarantee scheme independence
of the coefficients of the
singularities in the anomalous dimensions at least in the
nonsinglet case.
Indeed a resummation of these singularities in the nonsinglet anomalous
dimensions has been proposed long ago\ref\lipns{R.~Kirschner and L.~Lipatov,
\NP\vyp{B213}{1983}{122}.}. In the absence of appropriate
factorization theorems, it is unclear whether this
resummation, which
reproduces~\ref\blu{J.~Bl\"umlein and A.~Vogt, preprint DESY 95-175,
{\tt hep-ph/9510410}.} the known NLO leading singularity,
correctly reproduces the leading singularities
of the nonsinglet anomalous dimensions to all orders. If it does,
it implies
that the polarized and unpolarized nonsinglet quark distributions
will display the same small $x$--behavior, in contradiction
to Regge theory. More generally, one would
expect\ref\summ{R.~D.~Ball and S.~Forte, {\it Phys. Lett.}
{\bf B351} (1995) 313.} the summation of
logarithmic effects in $\ln{1\over x}$ induced by these singularities
to lead to a rise of parton distributions, which should be asymptotically
power-like; this rise should set in significantly faster than
 in the unpolarized case, due to the stronger nature of the singularity,
and thus, contrary to the unpolarized case\cite\summ, it could set
in before perturbation theory breaks down.

The combination of all these effects makes it rather hard to infer
the asymptotic small $x$ behavior of polarized parton distributions
from present-day data: a HERA-like kinematic coverage would be
necessary to disentangle the various effects. The only firm
prediction at this stage is that parton distributions at small $x$
will certainly grow (positive or negative) at least as fast
as eq.~\smxb, including the nonsinglet distributions, and that
therefore a simple phenomenological extrapolation based on Regge
expectations is not adequate at least in principle. More
detailed statements can be made by studying the structure
of the data in the $(x,Q^2)$ plane.
\medskip
\subsec{The evolution of $g_1(x,Q^2)$}
\smallskip
The only way of arriving at a precise and reliable determination of
the moments of $g_1$ is to describe its evolution in the $(x,Q^2)$
plane in terms of the evolution of polarized parton distribution. The
fact that these evolution effects are large not only
implies that they must be included in the data analysis, but also that
they might provide useful information on the polarized parton
content of the nucleon, and specifically on its gluon content. In
fact, one would expect $g_1$ to be a very sensitive probe
of $\Delta g$ --- much more than the unpolarized structure function
$F_2$ is a probe of $g$ --- since, even though the coupling of the gluon
to $g_1$ is formally NLO, thanks to the anomaly it does not decouple
asymptotically\cite\alros. In comparison to the unpolarized case, however,
the analysis is complicated by the fact that the singlet and nonsinglet
display a similar perturbative behavior in the small $x$ region: this
implies that only combining proton and neutron (or deuteron) data
is it possible to disentangle the contributions of the various
parton distributions to $g_1$, at least using data with the current
relatively restricted kinematic coverage.

Thanks to the recent determination\cite\merner of the full matrix
of two-loop anomalous dimensions, it is possible to study the evolution
of parton distributions at NLO. The computations
of ref.~\cite\merner\ are performed in the \MS\ scheme, where the
gluon decouples from the first moment of $g_1$ and thus is not properly
factorized, as we discussed in sect.~3.2. However, since the quark
and gluon coefficient
functions are known\refs{\koda,\alros}
in schemes where the gluon does contribute to $g_1$, the corresponding
anomalous dimensions may be determined\cite\BFRb\ from those
of ref.~\cite\merner.
A determination of parton distributions can then be made by
fitting to the data, provided one makes some simplifying assumptions on
the form of the starting parton
distributions.\footnote*{\footnotefont\baselineskip6 pt For a
review of LO polarized parametrizations see
ref.~\ref\ppar{T.~Gehrmann and
W.~J.~Stirling, Durham preprint DTP/95/78, {\tt hep-ph/9510243}.}.
A NLO fit has been presented in
ref.~\ref\glure{M.~Gl\"uck et al., Dortmund preprint DO-TH 95-13, {\tt
hep-ph/9508347}.}, but in a scheme where the gluon does not couple to $g_1$.
Approximate NLO fits (prior to the full determination of NLO anomalous
dimensions) in properly factorized schemes are in
refs.~\nref\stirgeold{T.~Gehrmann and
W.~J.~Stirling \ZP \vyp {C65}{1995}{461}.}\refs{\stirgeold,\BFR}.
A full NLO computation in
a variety of physical schemes is in ref.~\cite\BFRb.}
Specifically,
 a simple polynomial parametrization
of parton distributions may be adopted,
which entails the assumption that the asymptotic small $x$ behavior of parton
distributions starts setting in at the values of $x$
explored at present:
\eqn\parm{\Delta f(x, Q_0^2)=N\left(\alpha_f,
\beta_f,a_f\right)
\eta_fx^{\alpha_f}
(1-x)^{\beta_f}(1+a_fx)}
(where  $N(\alpha,\beta,a)$ is a normalization factor, fixed e.g. as
$N(\alpha,\beta,a)\int_0^1\!dx\, x^{\alpha}
(1-x)^{\beta}(1+ax)=1$).

Two fits\cite\BFR\ to the proton data only of parton distributions
of this form  are displayed in fig.~3,
together with the corresponding predicted deuteron
structure function. The nonsinglet
is assumed to be valence-like ($\alpha_{\rm NS}=+0.2$); the
gluon is assumed to be flat ($\alpha_g=0$) and the singlet quark  is
either steep ($\alpha_q=-0.5$, fig.~3i) or flat ($\alpha_q=0$, fig.~3ii).
The starting gluon is fixed either by requiring $\Delta \Sigma(1)$
to satisfy the Zweig rule, so that the fit then forces a
large gluon component to compensate (``maximal gluon'', fig.~3i),
or by symply taking the gluon distribution to vanish at the
starting scale (``minimal gluon'', fig~3ii). The
quality of these
two fits does not differ in a statistically significant way:
while a larger gluon corresponds to stronger evolution effects,
the proton data alone do not allow us to fix its size, or the small $x$
behavior of the various parton distributions.
Notice, however, that in order to obtain a satisfactory description
of the data it is necessary to introduce a gluon coupling
to $g_1$. This is due to the
fact that the SMC and E143 experiments measure $g_1$
 at different values of $Q^2$
for equal values of $x$. The two data sets can then
be made consistent with each other only if there is a sufficient
amount of perturbative evolution, and  this can hardly be
obtained in a LO computation\ref\anrev{G.~Altarelli, P.~Nason and
G.~Ridolfi, {\it Phys. Lett.} {\bf B320} (1994) 152},
or if schemes where the gluon decouples from
$g_1$ are adopted.

\topinsert
\vskip-3.5truecm\vbox{\hbox{\hskip-1truecm
\hfil\epsfxsize=10truecm\epsfbox{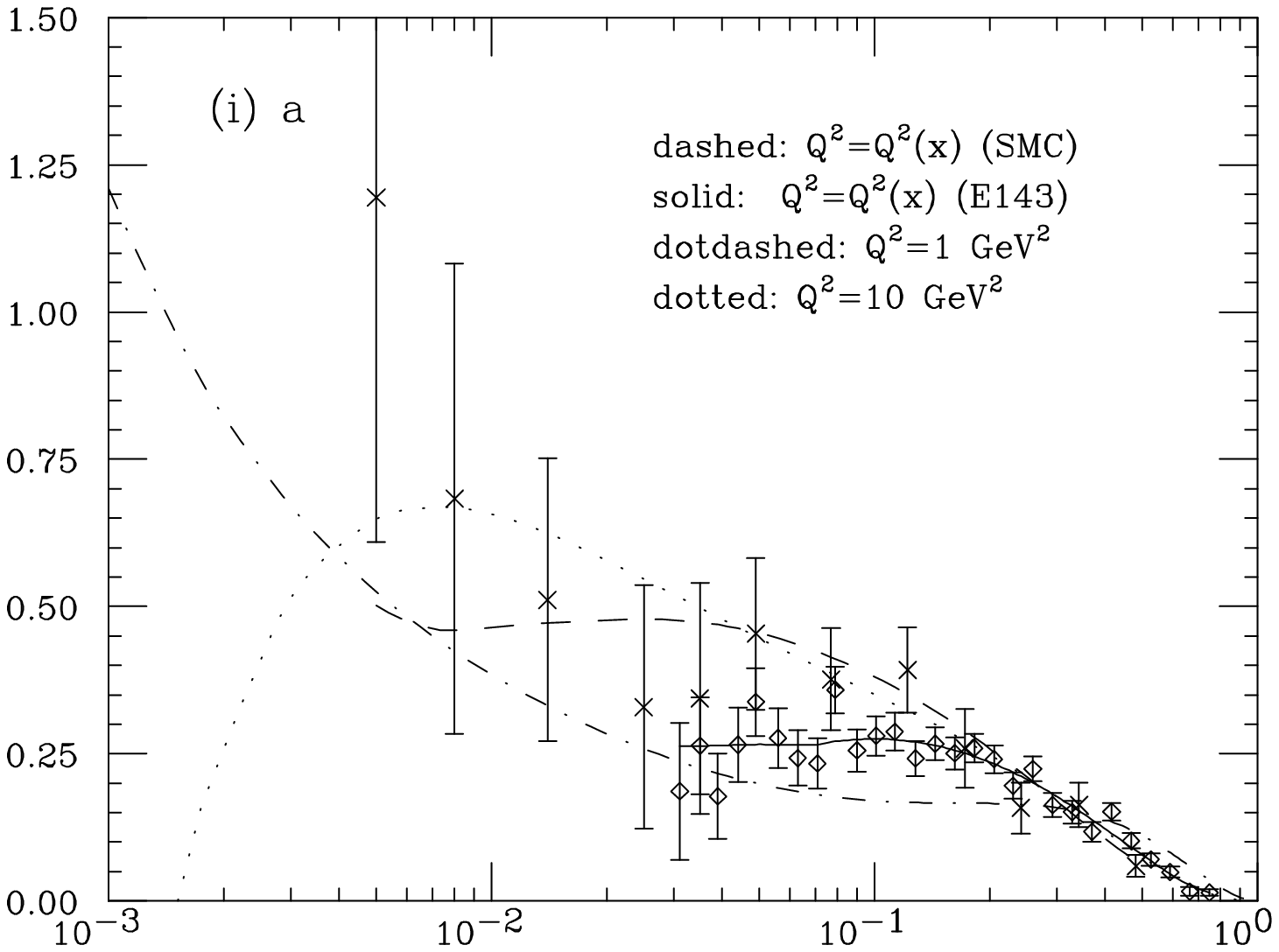}\hskip-2.truecm
\epsfxsize=10truecm\epsfbox{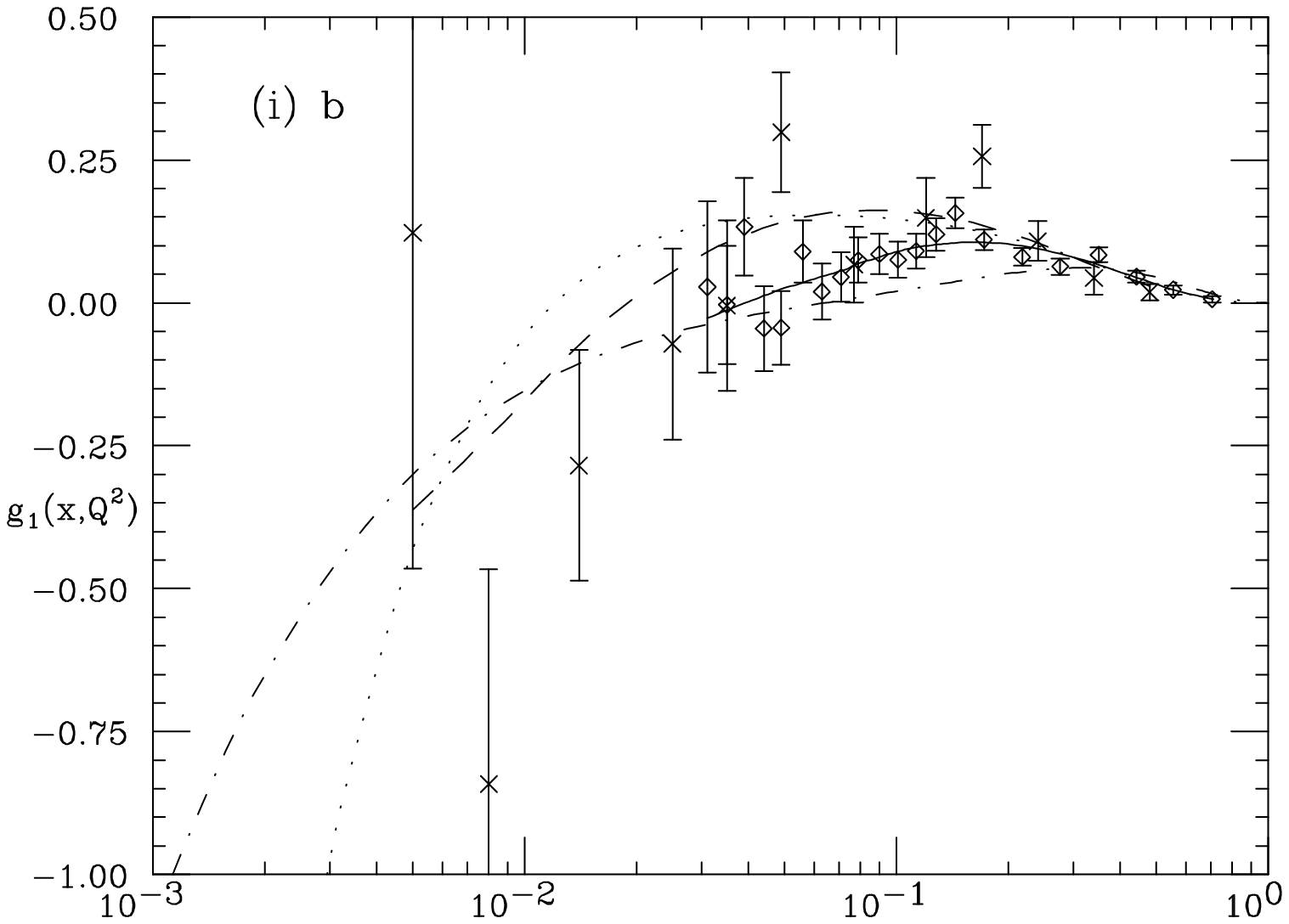}\hfil}
\vskip-7.5truecm
\hbox{\hskip-1truecm
\hfil\epsfxsize=10truecm\epsfbox{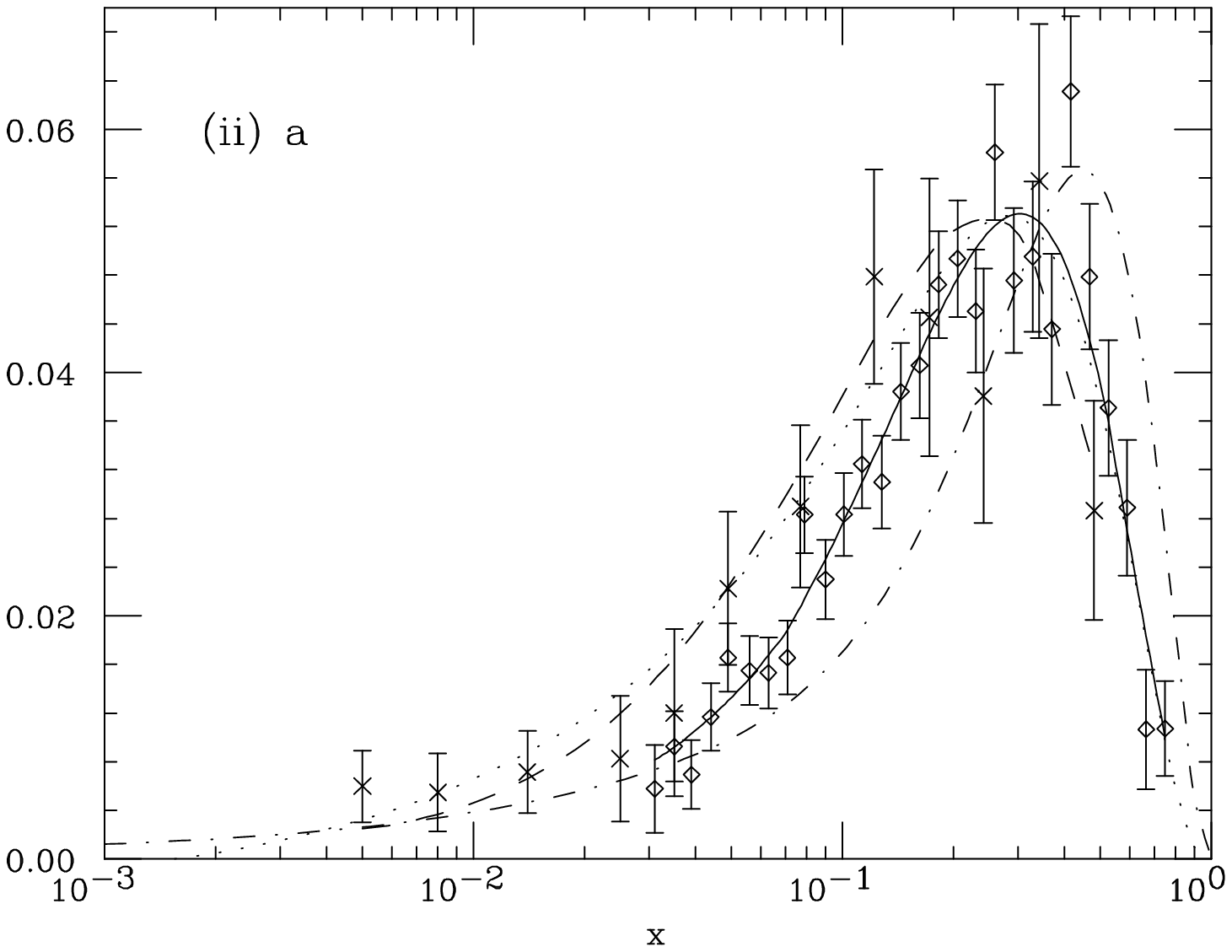}\hskip-2.truecm
\epsfxsize=10truecm\epsfbox{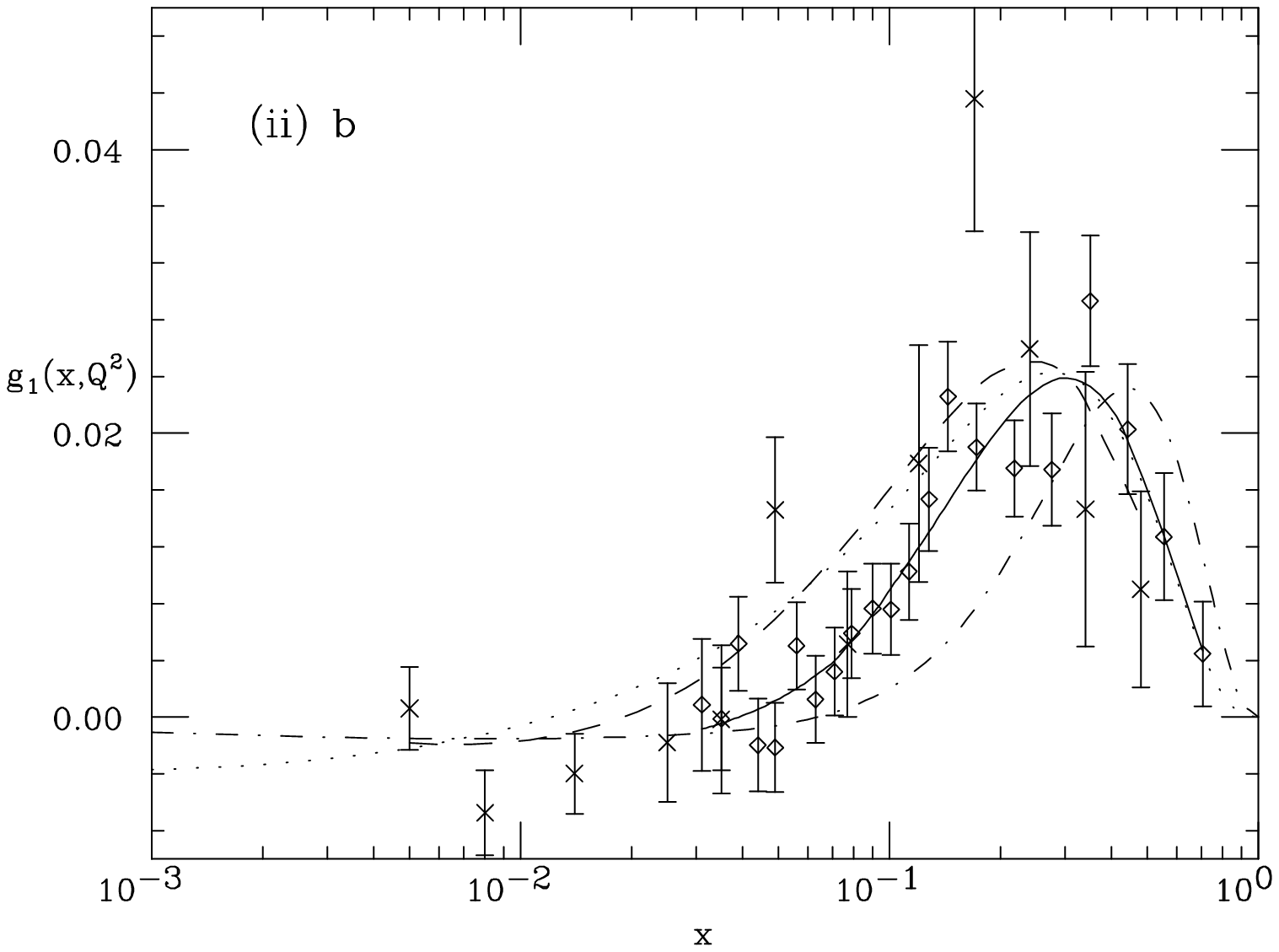}\hfil}
\vskip-4.truecm
\bigskip\noindent{\abstractfont\baselineskip6pt\narrower
Figure 4: Plot of  (i) $g_1(x)$ and (ii) $xg_1(x)$
for (a) proton and (b) deuteron.
The crosses are SMC data and the diamonds E143 data.
The curves correspond to a NLO fit\cite\BFRb
to all data.}}
\medskip
\endinsert
 Inspection of the
deuteron predictions (fig.~3b), however, clearly
shows  that more definite conclusions may be reached by  combining the
two sets of data: for instance the ``minimal gluon'' case seems to be
excluded. This is a consequence of the fact that the deuteron is a more
sensitive probe of the singlet and thus of the gluon, as explained in
sect.~3.1.
The results of a full NLO determination\cite\BFRb\ of $g_1$
(displayed in fig.~4)
indeed confirm that all the parameters
in eq.~\parm\ may then be determined with reasonable accuracy.
In particular, it is possible to disentangle the singlet
and nonsinglet contributions, and thus determine independently
the small $x$ behavior of the various parton distributions.
This
turns out to be steep for the nonsinglet
($\Delta q_{\rm NS}\neath\sim{x\to 0}x^{\alpha_{\rm NS}}$ with
$\alpha_{\rm NS}=-0.7\pm 0.2$), and valence-like for the
singlet (with large error). The result is quite suggestive in view of
the discussion in the previous section: it seems consistent
with the expectation, based on the resummations of
ref.~\cite\lipns,
that the polarized and unpolarized  nonsinglets should behave in
the same way, despite the prediction of Regge theory. However,
the asymptotic small $x$ behavior is only starting to set in at
the smallest values of $x$ covered by present--day data, which, in this
respect, are to be
compared with the NMC
unpolarized data~\cite\NMC: a kinematic
coverage comparable with that available at HERA for
determining~\cite\DAS\ the small $x$ behavior of $F_2$ would be
required to determine precisely the small $x$ behaviour
of polarized parton distributions.

\topinsert\hfil
\vbox{\tabskip=0pt \offinterlineskip
      \def\tablerule{\noalign{\hrule}}
      \halign to 400pt{\strut#&\vrule#\tabskip=1em plus2em
                   &\hfil#\hfil&\vrule#
                   &#\hfil&\vrule#
                   &#\hfil&\vrule#
                   &#\hfil&\vrule#
                   &#\hfil&\vrule#
                   &\hfil#&\vrule#\tabskip=0pt\cr\tablerule
      &&\omit\hidewidth $ Q^2 $ \hidewidth
      &&\omit\hidewidth $ \Delta \Sigma(N=1) $ \hidewidth
      &&\omit\hidewidth $ \Delta g (N=1) $ \hidewidth
      &&\omit\hidewidth $\Gamma_1^p$\hidewidth
      &&\omit\hidewidth $\Gamma_1^d$\hidewidth
             &&\omit\hidewidth $a_0$\hidewidth&\cr\tablerule
    && 3 &&$0.5\pm0.1$ &&$2.3\pm1.2$ &&$0.118\epm{0.016}{0.014}$ &&
         $0.022\epm{0.015}{0.013}$ && $0.15\epm{0.17}{0.12}$  &\cr\tablerule
    && 5 &&$0.5\pm0.1$ &&$2.6\pm1.4$ &&$0.120\epm{0.016}{0.014}$
      &&
         $0.023\epm{0.015}{0.013}$ && $0.14\epm{0.16}{0.11}$  &\cr\tablerule
    && 10 &&$0.5\pm0.1$ &&$3.0\pm{1.6}$ &&$0.122\epm{0.018}{0.014} $&&
         $0.023\epm{0.016}{0.013}$ && $0.14\epm{0.16}{0.11}$  &\cr\tablerule
}}
\hfil\bigskip\noindent{\abstractfont\baselineskip6pt\narrower
Table 2: Next-to-leading order determination\cite\BFRb\ of polarized first
moments. \smallskip}
\medskip
\endinsert
Present--day data are however sufficient to determine the singlet
first moments $\Delta \Sigma(1,Q^2)$ and $\Delta g(1,Q^2)$
with reasonable accuracy (see table~2), assuming
isospin [and SU(3)] to be exact. Even though
the first moment of $\Gamma_1$ at each $Q^2$
determines  only the linear combination eq.~\singao\
of $\Delta \Sigma(1,Q^2)$ and $\Delta g(1,Q^2)$,
their
different scale dependence is sufficient to determine them
independently: the large evolution
effect already observed
requires a large gluon component. Equation~\singao\ then forces a large value
of $\Delta \Sigma(1,Q^2)$,
 consistent with the Zweig rule. Notice that the direct gluon contribution
to $g_1$ is essential in order to get good agreement with the data.
Thus current data support the
proposal of ref.~\cite\alros\
that the smallness of  $a_0$ is due to a large gluon contribution to it.

The first moment of the quark and gluon distribution then determine
both the axial charge and the first moment of $\Gamma_1$. These determinations
turn out to be significantly smaller, and with significantly larger
error, than the corresponding determinations
from the experimental collaborations (table~1):
as discussed in sect.~3.1, a difference of the order of 10\% in the
value of $\Gamma_1^p$ results in a variation  of a factor of about
2 in the value of $a_0$. We may actually understand in detail the origin
of this discrepancy\cite\BFR, which is due to the approximate
treatment of  both  scale dependence
and extrapolation at small $x$ used in refs.\refs{\smcp-\slacd}.
Firstly, the data points which provide the bulk of $\Gamma_1$ are taken at a
scale $Q^2$ larger than the nominal average scale of either experiment: for
example in refs.\refs{\smcp,\smcd}
$Q^2_{\rm exp}(x)$ is always larger than $20$~GeV$^2$ for
$x$ above $0.1$,  while $\langle Q^2\rangle=10$~GeV$^2$ for this experiment.
Since in this region $g_1$ increases as the scale increases
the substantial
underestimate of evolution effects due to their approximate treatment in
refs.\refs{\smcp-\slacd} leads to an
overestimate of $\Gamma_1$. Furthermore, the flat small $x$ extrapolation
based on Regge theory
overestimates the corresponding contribution,
because it is obtained by extrapolating the
smallest--$x$ data points, which are taken at very low $Q^2$ ($\sim 1$~GeV$^2$)
but assumed to apply at larger values of $Q^2$
where perturbative evolution, as explained in sect.~4.1, leads
rapidly to negative values of $g_1$ and thus to a small (or even negative)
contribution to $\Gamma_1$ from the small $x$ region.

The presence of large corrections due to perturbative evolution then
inevitably implies sizable uncertainties related to
insufficiently precise knowledge of the polarized
gluon distribution, which drives this evolution
at least at the level of the first moment, and to the unknown
higher  order corrections (scheme ambiguities). These are indeed
the dominant sources
of the error on $\Gamma_1$ and thus $a_0$ in table~2. It is interesting
to notice that even though the values of table~2 are obtained by assuming
isospin [i.e. the Bjorken sum rule eq.~\bjsr] to be exact,
the values of
$a_0$, $\Delta \Sigma(1)$ and $\Delta g(1,Q^2)$ are essentially
insensitive to SU(2) [or SU(3)] violations. This suggests that
if the isotriplet first moment $\Gamma_1$ were
also determined by a global NLO analysis, rather than being fixed
using the Bjorken sum rule, its value might not differ much
from that found by combining  values of $\Gamma_1$ from table~1.
Such an analysis would however be required in order to test
the Bjorken sum rule in a reliable way.

\bigskip
\newsec{Outlook}
\medskip
Polarized structure functions determined
in deep-inelastic scattering are clearly the preferred
source of information on the polarized structure of the nucleon
at high energy, since the availability of renormalization
group methods and factorization theorems that characterizes fully
inclusive processes allows a determination of physical
observables which is free from the ambiguities due to our
ignorance of the low--energy dynamics. In particular,
due to the axial anomaly, polarized deep-inelastic
scattering provides an ideal probe of the polarized gluon distribution
and allows us to get a handle on quantities that have direct relevance
for our understanding of the chiral structure of the QCD vacuum.

The new generation of experiments\refs{\smcp-\slacd} has
allowed us to confirm these expectations, by providing us
with surprisingly accurate information on polarized parton
distributions. The analysis of these data, however,
  requires sophisticated theoretical and
phenomenological tools. A future generation of experiments
with a wider kinematic coverage in $x$ and $Q^2$, such as
those that would be possible with a polarized proton beam at
HERA\ref\ver{V.~Hughes, talk at the Workshop on Deep-Inelastic Scattering
and QCD, Paris, May 1995.}, could lead to a full understanding of
the small $x$ behavior and scale dependence of polarized parton distributions,
thereby testing our understanding of QCD in perturbation theory and
beyond.

\smallskip
\nobreak
\noindent{\bf Acknowledgements} I thank for several
discussions G.~Altarelli,
G.~Ridolfi  and R.~Ball, whom I also thank for a critical reading of the
manuscript.

\goodbreak

\medskip

\immediate\closeout\rfile\writestoppt
\baselineskip=10pt\footnotefont
\bigskip{\frenchspacing%
\parindent=20pt\escapechar=` \input refs.tmp\vfill\eject}\nonfrenchspacing
\vfill
\eject
\bye